\documentclass[%
 aip,
 amsmath,amssymb,
 reprint,%
]{revtex4-2}

\usepackage{graphicx}
\usepackage{dcolumn}
\usepackage{bm}

\usepackage[utf8]{inputenc}
\usepackage[T1]{fontenc}
\usepackage{mathptmx}
\usepackage{etoolbox}

\makeatletter
\def\@email#1#2{%
 \endgroup
 \patchcmd{\titleblock@produce}
  {\frontmatter@RRAPformat}
  {\frontmatter@RRAPformat{\produce@RRAP{*#1\href{mailto:#2}{#2}}}\frontmatter@RRAPformat}
  {}{}
}%
\makeatother
\begin{document}

\preprint{AIP/123-QED}

\title[Improvement of the Simmons model for tunnel junctions]{Improvement of the Simmons model for tunnel junctions}

\author{I. M. W. Räisänen}
\author{I. J. Maasilta}
\affiliation{Nanoscience Center, Department of Physics, University of Jyv\"askyl\"a, P.O. Box 35, FI-40014 Jyv\"askyl\"a, Finland}
\email{ilmo.m.w.raisanen@jyu.fi}
\email{maasilta@jyu.fi}

\date{\today}

\begin{abstract}
The Simmons model is a well-known and widely used model for the elastic tunneling current of a metallic tunnel junction, and fitting it to electrical measurements can be used to estimate thicknesses and heights of the tunnel barriers. We present here an improvement of the Simmons model, deriving new more accurate analytical formulas for the tunneling current density and conductance at finite voltage and temperature. We demonstrate that our conductance-voltage formulas are much closer to the Wentzel-Kramers-Brillouin approximation than the Simmons model and its commonly used simplified parabolic approximation. In addition, we demonstrate the practical use of our model, by fitting it to experimental tunnel junction conductance-voltage data and showing a sizeable difference from the Simmons model.
\end{abstract}

\maketitle

\section{Introduction} \label{Introduction}

Tunnel junctions\cite{Wolf}, i.e., two electrical conductors separated by a thin electrically insulating barrier allowing quantum mechanical electron tunneling, are versatile and highly useful devices in many application areas of physics. In particular, superconducting tunnel junctions have in recent years been widely applied as components of various advanced devices, such as building blocks of superconducting quantum bits \cite{qubits}, superconducting quantum interference devices (SQUIDs) \cite{SQUIDbook}, sensitive radiation detectors \cite{Enss}, on-chip electronic coolers and thermometers \cite{Giazotto,Muhonen} and metrological applications for the definition of Ampere\cite{Pekola}. This pull from applications has generated renewed interest from the materials science perspective to explore more advanced barrier materials and fabrication methods \cite{Murray2021,akipaper}, in addition to the most commonly used thermally grown amorphous aluminum oxide barriers \cite{Wolf}. 

With such experimental advances, it is also important to critically review established theories, as they describe how various material parameters affect the tunneling current. We limit the discussion here to the case of metal-insulator-metal tunnel junctions, for the sake of simplicity. From the barrier perspective, junctions with one or two superconducting electrodes can always be measured as normal metal junctions above the superconducting critical temperature, to avoid the strong extra features in the tunneling current originating from superconductivity \cite{Wolf}. In most cases, it is enough to work with the Wentzel-Kramers-Brillouin (WKB) approximation \cite{Wolf,Brinkman}, which, however, requires numerical integration even for the simplest barrier shapes, and is therefore not very practical when trying to determine the barrier properties from experimental data. For that reason, approximate analytical formulas 
describing the elastic tunneling current as a function of bias voltage and temperature were developed in the past, most notably by Simmons\cite{SimmonsLowVoltage, SimmonsSimilarElectrodes, SimmonsDissimilarElectrodes, SimmonsThermal}. His formulas were derived for a general barrier shape, the main parameters being the barrier thickness and the average barrier height. 

The results by Simmons demonstrate that the barrier parameters not only influence the value of the zero-bias tunneling conductance, but also the rate at which the junction conductance increases with bias voltage and temperature. In other words, the tunneling current is not a purely linear (Ohmic) function of the voltage. This allows fitting of the conductance-voltage $(G-V)$ curves to determine both barrier height and thickness independently. For that reason, Simmons' model and its simplified low-voltage parabolic conductance-voltage $(G-V)$ expansion have been widely used to estimate thicknesses and heights of tunnel barriers \cite{Koberidze, Galceran2015, Koppinen2007, Parkin2004, Gloos2003,  Dorneles2003, Rippard, Buchanan2002ApplPhysLett, Wang2001, Barner1989, Braginski1986}. Another widely used approximative analytical model for zero-temperature conductance was derived by Brinkman, Dynes and Rowell (BDR) \cite{Brinkman} for the case of an asymmetric trapezoidal barrier, giving identical results to the Simmons model for the curvature and the value of the conductance minimum in the lowest order.    

In this paper, we revisit the theory for metal-insulator-metal tunnel junctions within the WKB approximation and derive new, more accurate analytical expressions for tunneling current and conductance as a function of both bias voltage and temperature. We demonstrate that our expressions approximate the numerical WKB results much more closely than the old expressions in the literature, using realistic barrier parameters. We also derive expressions for the low-voltage parabolic approximation for the conductance-voltage $G(V)$ curves within our model, including the temperature dependence, which can easily be used to fit experimental data. As an example, we fit a set of experimental $G-V$ data, demonstrating a sizeable difference to the widely used parabolic approximation of the Simmons model, even for the most common barrier material of thermally grown aluminum oxide.

\section{Tunneling in the WKB approximation}

\begin{figure}
\includegraphics[width=0.98\linewidth]{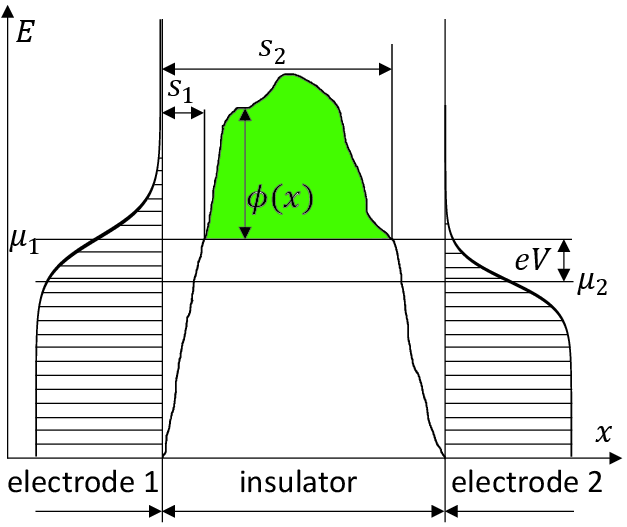} 
\caption{\label{GeneralizedTunnelBarrier} A general tunnel barrier formed by an insulating film between two metal electrodes. Electrode 1 with chemical potential $\mu_1=\mu$ is negatively biased with voltage $V$ with respect to electrode 2 that has a chemical potential $\mu_2=\mu-eV$ shifted by the bias voltage $V$. The net flow of electrons takes place from electrode 1 to electrode 2.}
\end{figure}

Let us consider a tunnel junction with a general tunnel barrier, illustrated in Fig.~\ref{GeneralizedTunnelBarrier}, with a barrier $\phi(x)$  depending on the position $x$ along the direction of current flow.  Here $\phi(x)$ is measured from the chemical potential $\mu$ of the negatively biased electrode (left electrode in Fig. \ref{GeneralizedTunnelBarrier}). According to the WKB approximation, the probability that an electron whose kinetic energy in the $x$ direction is $E_x$ can tunnel through the barrier is given by \cite{Wolf}
\begin{equation}
D(E_x) = \exp\left[-\frac{4\pi\sqrt{2m{^*}}}{h}\int_{x_1}^{x_2}(\mu+\phi(x)-E_x)^{\frac{1}{2}}dx \right],
\label{D(Ex)SecondForm}
\end{equation}
where $h$ is the Planck constant and $m^*$ is the effective mass of an electron in the barrier. The integration limits $x_1$ and $x_2$ are the classical turning points at energy $E_x$, which in general are thus functions of $E_x$ for a potential that does not have infinitely sharp edges. This energy dependence of the tunneling path length $x_2-x_1$ is typically not taken into account, and Simmons and others\cite{SimmonsSimilarElectrodes,Wolf} approximate the turning points to be constants $s_1$ and $s_2$, defined as the turning points at energy $\mu$ (Fig.~\ref{GeneralizedTunnelBarrier}). This approximation is exact for the case of rectangular and trapezoidal barriers, for which the edges are sharply defined and therefore the classical turning points are constants. 


With $D(E_x)$ known, the elastic tunneling current density $J$ as a function of voltage $V$ and temperature $T$ can then be calculated from \cite{SimmonsSimilarElectrodes,Wolf}

\begin{equation}
J(V,T) = B\int_{0}^{E_m}D(E_x)\int_{0}^{\infty} [f(E)-f(E+eV)]dE_rdE_x,
\label{MostGeneralJ}
\end{equation}
where the constant
\begin{equation}
B=\frac{4\pi m_{\textrm{e}}e}{h^3}, \nonumber
\label{B'}
\end{equation}
and where $m_{\textrm{e}}$ is the mass of the electron in the electrodes, taken here to equal the free-electron mass, $e$ is the elementary charge, $E_m$ is the maximum energy of the tunneling electrons, $f(E)$ and $f(E+eV)$ are the Fermi-Dirac distribution functions in electrodes 1 and 2, respectively, 
and $E_r=E_y+E_z$ is the transverse part of the total energy $E=E_x+E_y+E_z$.

At zero temperature, Eq. \eqref{MostGeneralJ} simplifies to \cite{Wolf, SimmonsSimilarElectrodes} 
\begin{gather}
J = B\Big[eV\int_{0}^{\mu-eV} D(E_x)dE_x
+\int_{\mu-eV}^{\mu}(\mu-E_x)D(E_x)dE_x\Big], 
\label{WKBJ(V)Integral}
\end{gather}
whereas at nonzero temperatures, the transverse energy integral in Eq. \eqref{MostGeneralJ} has an analytical form, giving the general solution\cite{Wolf, SimmonsThermal} 
\begin{gather}
J(V,T) = Bk_{\textrm{B}}T\int_{0}^{E_m} D(E_x)\nonumber\\
\times \ln \Bigg\{\frac{1+\exp[(\mu-E_x)/(k_{\textrm{B}}T)]}{1+\exp[(\mu-E_x-eV)/(k_{\textrm{B}}T)]}\Bigg\}dE_x,
\label{WKBJ(V,T)Integral}
\end{gather}
where $k_{\textrm{B}}$ is the Boltzmann constant.

\section{Foundations and main results of the Simmons model} \label{Foundations of Simmons' model}

Next, we critically review the results of Simmons' work. This is important as our modeling uses the same foundations, but extends the range of applicability of the original model. To be able to derive an analytical result for a general barrier shape, Simmons makes the key approximation to estimate the barrier shape function $\phi(x,V)$ by its average
\begin{equation}
\overline{\phi} = \frac{1}{\Delta s}\int_{s_1}^{s_2}\phi(x,V)dx, \nonumber
\label{overline_phi}
\end{equation}
where $\Delta s=s_2-s_1$ is the electrical thickness of the barrier at $\mu$ (Fig.~\ref{GeneralizedTunnelBarrier}), and we have made the dependence of the barrier shape on the bias voltage $V$ explicit. By expanding the integrand of Eq. \eqref{D(Ex)SecondForm} in the lowest order (See Appendix \ref{SimmonsJVGV} for details),  
Simmons derives Eq. (10) of Ref. \onlinecite{SimmonsSimilarElectrodes}, giving 
\begin{equation}
D(E_x) \simeq \exp\left[-A(\mu+\overline{\phi}-E_x)^{\frac{1}{2}}\right],
\label{D(Ex)}
\end{equation}
with
\begin{equation}
A = \frac{4\pi\beta \Delta s\sqrt{2m^{*}}}{h}, 
\label{A}
\end{equation}
where Simmons has defined $\beta$ as the lowest-order correction factor (Eq. (A6) of Ref. \onlinecite{SimmonsSimilarElectrodes}), which contains the dependence on the barrier shape: 
\begin{equation}
\beta = 1 - \frac{1}{8(\mu+\overline{\phi}-E_x)^2\Delta s}\int_{s_1}^{s_2}[\phi(x)-\overline{\phi}]^2dx. \nonumber
\label{beta}
\end{equation}
As pointed out by Hartman\cite{Hartman}, this factor is still a function of $E_x$, which prevents the analytical integration of Eq. \eqref{WKBJ(V)Integral}. Therefore, Simmons makes a further simplification that $\beta$ is approximately constant, which allows the derivation, along with various approximations (see details in Appendix \ref{SimmonsJVGV}), of a simple formula for current density as a function of voltage in the zero-temperature limit, Eq. (20) of Ref. \onlinecite{SimmonsSimilarElectrodes}:
\begin{equation}
J(V,0) = \frac{4B}{A^2}\Big\{\overline{\phi}\exp(-A{\overline{\phi}}^{\frac{1}{2}})
-(\overline{\phi}+eV)\exp[-A({\overline{\phi}}+eV)^{\frac{1}{2}}]\Big\}.
\label{SimmonsGeneralizedJV}
\end{equation}
Note that the above general result also contains an implicit voltage dependence through the voltage-dependent average barrier height $\overline{\phi} = \overline{\phi}(V)$, so that the actual calculation of the current-voltage characteristics using Eq. \eqref{SimmonsGeneralizedJV} requires the knowledge of the structure of the barrier. 

Considering nonzero temperatures, from Eq. \eqref{WKBJ(V,T)Integral} Simmons  also derives  an approximate formula for current density at finite voltage and temperature (see details in Appendix \ref{SimmonsJTGT}), which in the lowest order gives Eq. (11) of Ref. \onlinecite{SimmonsThermal}:
\begin{equation}
J(V,T) = J(V,0)\Big[1+\frac{(\pi Ak_{\text{B}}T)^2}{24\overline{\phi}(V)}\Big],
\label{SimmonsGeneralizedJ(V,T)}
\end{equation}
where we stress that the voltage dependence appears not only in the zero-temperature current density $J(V,0)$, but also in the temperature-dependent factor through the voltage-dependent average barrier height $\overline{\phi}(V)$, giving a second term for the finite-temperature conductance.  

\subsection{Trapezoidal barrier shape}
\begin{figure}[b]
\includegraphics[width=0.98\linewidth]{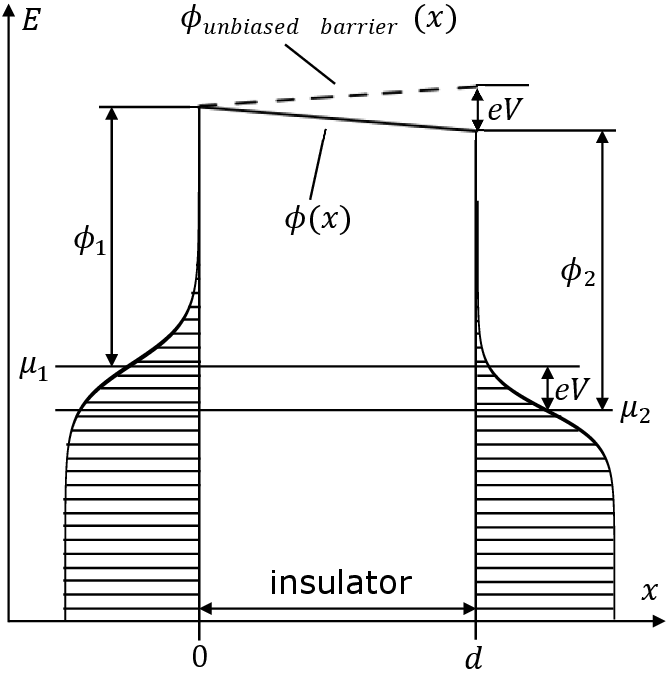} 
\caption{\label{BarrierBetweenDissimilarElectrodesFinished} An unbiased trapezoidal tunnel barrier (dashed line) and a voltage-biased barrier (solid line) with well-defined electrode-insulator interfaces and physical thickness $d_\textrm{phys} = d$. Electrons tunnel in $x$ direction.}
\end{figure}

As noted above, Simmons' results for the general barrier do not allow for an explicit calculation of the tunneling current-voltage characteristics without the knowledge of the shape of the barrier and its voltage dependence. For that reason, Simmons develops his results further for simplified models of rectangular and trapezoidal barrier shapes, which is an assumption made in the analytical BDR model\cite{Brinkman}, as well. Such barriers have well-defined electrically sharp electrode-insulator interfaces, as shown in Fig. \ref{BarrierBetweenDissimilarElectrodesFinished}, making the approximation of a constant tunneling path length $\Delta s = s_2-s_1 = d_\textrm{phys}$ exact for voltages lower than the barrier height. Let us call the zero-bias barrier heights at the two electrode-insulator interfaces $\phi_1$ and $\phi_2$ with $\phi_1 \leq \phi_2$. Assuming the barrier material has uniform dielectric properties, the applied voltage drops linearly across the barrier such that 
\begin{equation}
\phi(x) = \phi_1+\frac{x}{d_\textrm{phys}}(\phi_2-eV-\phi_1) 
\label{phi(x)}
\end{equation}
with $|V| \leq \phi_1/e$. Then $\beta=1$ to a good approximation \cite{SimmonsSimilarElectrodes, SimmonsDissimilarElectrodes}, and the average barrier height and the electrical thickness $\Delta s$ are given by
\begin{equation}
\left\{ \begin{array}{ll}
\overline{\phi} = \phi_0 - \frac{eV}{2}\\
\Delta s = d_\textrm{phys}
\label{overline_phi2}
\end{array}\right.
\end{equation}
where $d_\textrm{phys}$ is the physical thickness of the insulating film and $\phi_0=(\phi_1+\phi_2)/2$ is the average height of the unbiased barrier.

We should note here that Simmons uses in his equations the free-electron mass $m_{\textrm{e}}$ instead of the barrier effective band mass $m^*$ in the barrier penetration factor $A$ defined in Eq. \eqref{A}. From Eq. \eqref{A}, we see that such a substitution corresponds to the barrier thickness in Simmons' notation being an effective thickness $d_\textrm{eff}$, which is related to the physical thickness $d_\textrm{phys}$ by the ratio of the two masses as 
\begin{equation}
d_\textrm{eff} = d_\textrm{phys}\sqrt{\frac{m^{*}}{m_\textrm{e}}}.
\label{dphys}
\end{equation}
For the sake of simplicity of notation and to allow direct comparison to Simmons' equations, for the rest of this article we also write equations using only $m_{\textrm{e}}$, so that $d$ always denotes the effective thickness $d \equiv d_\textrm{eff}$ instead of the physical one.



Instead of current, experiments often measure the differential conductance $G = dI/dV$. By substituting Eq. \eqref{overline_phi2} into Eq. \eqref{SimmonsGeneralizedJV} and differentiating with respect to voltage (see Appendix \ref{SimmonsJVGV}), one can derive an equation for the conductance $G$ per junction area $A_\textrm{j}$ in the low-temperature limit
\begin{gather}
\frac{G}{A_\textrm{j}} = \frac{e^2}{8\pi hd^{2}}\Bigg\{\exp\left[-A\left(\phi_0-\frac{eV}{2}\right)^{\frac{1}{2}}\right]\left[A\left(\phi_0-\frac{eV}{2}\right)^{\frac{1}{2}}-2\right]\nonumber\\
+ \exp\left[-A\left(\phi_0+\frac{eV}{2}\right)^{\frac{1}{2}}\right]\left[A\left(\phi_0+\frac{eV}{2}\right)^{\frac{1}{2}}-2\right]\Bigg\},
\label{SimmonsGV}
\end{gather}
where $A = 4\pi d\sqrt{2m_\textrm{e}}/h =  2d\sqrt{2m_\textrm{e}}/\hbar$. Eq. \eqref{SimmonsGV} does not appear in Simmons' papers, and we are not aware of it appearing in any previous publications. However, we still call Eq. \eqref{SimmonsGV} Simmons' $G-V$ equation, as its derivation from Simmons' main result, Eq.  \eqref{SimmonsGeneralizedJV}, is so straightforward. It is \emph{not} one of the main results of this work, as we derive a more accurate $G-V$ formula in Section \ref{newresults} below.

A more widely used low-temperature $G-V$ formula based on Simmons' equations can be derived around zero bias in the small-bias limit $eV \ll \phi_0$ by Taylor expanding Eq. \eqref{SimmonsGV} around $V=0$ to second order.  By neglecting smaller terms (see Appendix \ref{SimmonsJVGV}), it is possible to write the lowest-order result as
\begin{equation}
G(V,T=0) = G_{0,0}\left(1+\frac{V^2}{V_{0,0}^2}\right),
\label{SimmonsSimpleGVatT=0}
\end{equation}
with
\begin{equation} 
G_{0,0} = \frac{e^2A_\textrm{j}\sqrt{2m_\textrm{e}\phi_0}}{h^2d}\textrm{exp}\biggr[\frac{-2d\sqrt{2m_\textrm{e}\phi_0}}{\hbar}\biggr] 
\label{G_00}
\end{equation}
the lowest-order zero-bias conductance and
\begin{equation}
V_{0,0}^2 = \frac{4\hbar^2\phi_0}{e^2m_\textrm{e}d^2}
\label{V_00}
\end{equation}
a constant whose inverse is proportional to the (positive) curvature around zero bias. 
We call Eq. \eqref{SimmonsSimpleGVatT=0} Simmons' simplified parabolic $G-V$ approximation even though Simmons only shows equations for current density $J$ in his papers \cite{SimmonsLowVoltage, SimmonsSimilarElectrodes, SimmonsDissimilarElectrodes, SimmonsThermal}. Eq. \eqref{SimmonsSimpleGVatT=0} has been mentioned \cite{Gloos2003, Koppinen2007, Feshchenko2017} and explicitly used in studies of properties of tunnel barriers \cite{Gloos2003, Koppinen2007}. We also question the accuracy of Eq. \eqref{SimmonsSimpleGVatT=0} and introduce an improved parabolic low-bias approximation in Sect. \ref{newresults} below.  

We should point out that both Eqs. \eqref{SimmonsGV} and \eqref{SimmonsSimpleGVatT=0} show that the conductance in Simmons' models is symmetric with bias polarity, even for the case of an asymmetric trapezoidal barrier with $\phi_1 \neq \phi_2$. This follows from the assumption of a constant value for $\beta$, which physically corresponds to approximating the shape of the zero-bias barrier as a rectangle with an average height $\phi_0$. This simplification washes out the asymmetry with bias polarity in conductance-voltage characteristics due to the asymmetric trapezoidal shape\cite{Brinkman,Hartman}, whose main effect is to shift the conductance minimum from $V=0$ to a finite value\cite{Brinkman}. Thus, models based on Simmons' approach can still be applied to asymmetric data by shifting the minimum of the data to $V=0$.       

As mentioned above, Simmons also derived a quadratic correction term for $J(V)$ due to finite temperatures, Eq. \eqref{SimmonsGeneralizedJ(V,T)}. By differentiating w.r.t. $V$ and using the same approximations as for Eq. \eqref{SimmonsSimpleGVatT=0} (Appendix \ref{SimmonsJTGT}), it is also possible to write for the zero-bias conductance $G(0,T)$ a simple result for the trapezoidal barrier
\begin{equation}
G(V=0,T) = G_{0,0}\left(1+\frac{T^2}{T_{0,0}^2}\right)
\label{SimmonsG(T)a}
\end{equation}
with
\begin{equation}
T_{0,0}^2 = \frac{3\hbar^2\phi_0}{\pi^2k_{\textrm{B}}^2m_\textrm{e}d^2} 
\label{T_00}
\end{equation}
now a constant whose inverse is proportional to the (positive) curvature around zero temperature. Eq. \eqref{SimmonsG(T)a} has been presented and used in Ref. \onlinecite{Gloos2000}. 
The size of this finite-temperature correction is not very large: For typical barrier heights $\phi_0 \approx 1$ eV and thicknesses $d \approx 1$ nm, $T_{0,0} \approx 1800$ K, giving a 3 \% correction at room temperature and an insignificant factor $< 0.2$ \% already at 77 Kelvin. This means that zero-temperature $G(V)$ equations without temperature corrections can typically be used at cryogenic temperatures.   

To conclude the discussion on Simmons' models, we are left with two key questions: (i) Are all the approximations used to derive Simmons model Eqs. \eqref{SimmonsGeneralizedJV}, \eqref{SimmonsGeneralizedJ(V,T)}, \eqref{SimmonsGV}, \eqref{SimmonsSimpleGVatT=0} and \eqref{SimmonsG(T)a} justified for realistic barrier parameter values, and if not, is it possible to derive new, more accurate expressions? (ii) What is the lowest-order conductance model for trapezoidal barriers that takes simultaneously into account both the finite bias and finite temperature? We answer those questions below by deriving new analytical formulas that improve on the above results from the Simmons model.

\section{Improved analytical formulas for current density and conductance at finite voltage and temperature} \label{newresults}
In deriving Eq. \eqref{SimmonsGeneralizedJV}, 
Simmons makes approximations (see Appendix \ref{SimmonsJVGV}) that can lead even the generalized equation Eq. \eqref{SimmonsGV}  quite far away from the WKB approximation starting point, Eq. \eqref{WKBJ(V)Integral}, as we will later see for the example case of a rectangular barrier in Fig. \ref{ComparisonOfGVModels5CurvesShorterVRange}. Even more critically, there are mathematically non-rigorous steps in Simmons' derivation of the finite-temperature correction, Eq. \eqref{SimmonsGeneralizedJ(V,T)}, as we show in Appendix \ref{SimmonsJTGT}.  
To overcome these issues, we derive and present here new $J(V,T)$ and $G(V,T)$ equations, which are more accurate than all previous Simmons model equations, but still based on the fundamental approximation of Simmons, Eq. \eqref{D(Ex)}. Results are derived to improve both Simmons' generalized results for current density, Eqs. \eqref{SimmonsGeneralizedJV} and \eqref{SimmonsGeneralizedJ(V,T)}, and the trapezoidal model equations for conductance, Eqs. \eqref{SimmonsGV}, \eqref{SimmonsSimpleGVatT=0} and \eqref{SimmonsG(T)a}. In addition, we derive equations for conductance $G(V,T)$ as a function of both the voltage and temperature for the first time\footnote{Eq. (2) in Ref. \cite{Gloos2000} is not correct, as discussed in Appendix \ref{SimmonsJTGT}.}.

By substituting Eq. \eqref{D(Ex)} into Eq. \eqref{WKBJ(V,T)Integral}, integrating by parts twice and Taylor expanding the more slowly varying factor of the integrand in energy around $\mu$ and $\mu-eV$ to second order (Sommerfeld expansion, see details in Appendix \ref{J(V,T)}), we get
\begin{gather}
J(V,T) = \frac{4B}{A^4}\Bigg\{\exp(-A{\overline{\phi}}^{\frac{1}{2}})\Big[A^2\overline{\phi}+3(A{\overline{\phi}}^{\frac{1}{2}}+1)+\frac{A^4\pi^2k_{\textrm{B}}^2T^2}{24}\Big]\nonumber\\
-\exp\Big[-A(\overline{\phi}+eV)^{\frac{1}{2}}\Big]\Bigg[A^2(\overline{\phi}+eV)\nonumber\\
+3\Big[A(\overline{\phi}+eV)^{\frac{1}{2}}+1\Big]+\frac{A^4\pi^2k_{\textrm{B}}^2T^2}{24}\Bigg]\Bigg\},
\label{ImprovedGeneralizedJ(V,T)}
\end{gather}
where $A$, $B$ and $\overline{\phi}$ have the same definitions as before in Sect. \ref{Foundations of Simmons' model}. The expansion used in the derivation is accurate for temperatures $k_BT < \overline{\phi}$, which is easily satisfied even at room temperature for bias voltages $eV < \phi_0$.  

Eq. \eqref{ImprovedGeneralizedJ(V,T)} is a generalized formula that describes the current density across a tunnel barrier of arbitrary shape at finite voltage and temperature. In addition, we derive in Appendix \ref{J(V,0)} the zero-temperature version of Eq.  \eqref{ImprovedGeneralizedJ(V,T)}, $J(V,T=0)$, by starting directly from the general expression for the zero-temperature current density, Eq. \eqref{WKBJ(V)Integral}. The result is consistent with Eq. \eqref{ImprovedGeneralizedJ(V,T)} with the substitution $T=0$. Comparing to the Simmons model Eq. \eqref{SimmonsGeneralizedJV} at $T=0$, Eq. \eqref{ImprovedGeneralizedJ(V,T)} has the additional terms $3(A\overline{\phi}^{\frac{1}{2}}+1)\exp(-A\overline{\phi}^{\frac{1}{2}})$ 
and $-3[A(\overline{\phi}+eV)^{\frac{1}{2}}+1]\exp[-A(\overline{\phi}+eV)^{\frac{1}{2}}]$. At finite temperatures, unlike Simmons' Eq. \eqref{SimmonsGeneralizedJ(V,T)},  our improved Eq. \eqref{ImprovedGeneralizedJ(V,T)} cannot be written as a simple product of $J(V,0)$ and a temperature-dependent factor.

Next, we again consider the trapezoidal barrier shape with voltages less than the smaller barrier height $|V| < \phi_1/e$ (Fig. \ref{BarrierBetweenDissimilarElectrodesFinished}). By substituting Eq. \eqref{overline_phi2} into Eq. \eqref{ImprovedGeneralizedJ(V,T)} and differentiating with respect to voltage (see Appendix \ref{J(V,T)}) we get
\begin{gather}
\frac{G}{A_\textrm{j}} = \frac{e^2}{8\pi hd^{2}}\Bigg\{\exp\left[-A\left(\phi_0-\frac{eV}{2}\right)^{\frac{1}{2}}\right]\Bigg[A\left(\phi_0-\frac{eV}{2}\right)^{\frac{1}{2}}+1\nonumber\\
+\frac{A^3\pi^2k_{\textrm{B}}^2T^2}{24}\left(\phi_0-\frac{eV}{2}\right)^{-\frac{1}{2}}\Bigg]\nonumber\\
+\exp\left[-A\left(\phi_0+\frac{eV}{2}\right)^{\frac{1}{2}}\right]\Bigg[A\left(\phi_0+\frac{eV}{2}\right)^{\frac{1}{2}}+1\nonumber\\
+\frac{A^3\pi^2k_{\textrm{B}}^2T^2}{24}\left(\phi_0+\frac{eV}{2}\right)^{-\frac{1}{2}}\Bigg]\Bigg\}.
\label{ImprovedG(V,T)}
\end{gather}
Eq. \eqref{ImprovedG(V,T)} improves Simmons' $G-V$ Eq. \eqref{SimmonsGV} by two factors: It (i) includes the two additional temperature-dependent correction terms proportional to $T^2/(\phi_0-eV/2)^{\frac{1}{2}}$ and $T^2/(\phi_0+eV/2)^{\frac{1}{2}}$, and (ii) corrects the two constant terms from $-2$ to $+1$, which thus influence the results at $T=0$, as well.  
 
By writing the second-degree Taylor polynomial of Eq. \eqref{ImprovedG(V,T)} around $V=0$ for the low-bias limit $V \ll \phi_0/e$ (see Appendix \ref{J(V,T)}), we finally get a simple formula for the conductance as a function of both voltage and temperature, which is the main result of this work: 
\begin{gather}
G(V,T) = G_0\Bigg[1+\frac{4\pi^2k_{\textrm{B}}^2T^2}{3}(1+C)\left(\frac{1}{(eV_0)^2}+\frac{3}{32\phi_0^2C}\right)\Bigg]\frac{V^2}{V_0^2}\nonumber\\
+G_0\left(1+\frac{T^2}{T_0^2}\right),
\label{ParabolicApproximationOfImprovedG(V,T)}
\end{gather}
where
\begin{equation} 
G_0 = G_{0,0}(1+C), \nonumber
\label{G_0}
\end{equation}
\begin{equation}
V_0^2 = V_{0,0}^2(1+C), \nonumber
\label{V_0}
\end{equation}
and
\begin{equation}
T_0^2 = T_{0,0}^2(1+C) \nonumber
\label{T_0}
\end{equation}
are the new corrected coefficients over the old Simmons' coefficients $G_{0,0}$,  $V_{0,0}$, and $T_{0,0}$ defined by Eqs. \eqref{G_00}, \eqref{V_00} and \eqref{T_00}, respectively, and
\begin{equation}
C = \frac{\hbar}{2d\sqrt{2m_{\textrm{e}}\phi_0}}
\label{C}
\end{equation}
is the dimensionless correction factor.

Inspecting Eq. \eqref{ParabolicApproximationOfImprovedG(V,T)} further, we first see that at $T=0$, it reduces to Simmons model Eq. \eqref{SimmonsSimpleGVatT=0}, but with the corrected coefficients $G_0$ and $V_0$ instead of the "bare" coefficients $G_{0,0}$ and $V_{0,0}$, i.e.
\begin{equation}
G(V,T=0) = G_{0}\left(1+\frac{V^2}{V_{0}^2}\right).
\label{SimmonsG(V)}
\end{equation}
Similarly, at zero bias, Eq. \eqref{ParabolicApproximationOfImprovedG(V,T)} reduces to Simmons model Eq. \eqref{SimmonsG(T)a} with corrected coefficients $G_0$ and $T_0$ instead of the "bare" coefficients $G_{0,0}$ and $T_{0,0}$, i.e.
\begin{equation}
G(V=0,T) = G_{0}\left(1+\frac{T^2}{T_{0}^2}\right).
\label{SimmonsG(T)}
\end{equation}
Finally, and perhaps most interestingly, our model predicts that the curvatures of the $G(V)$ parabolae are not temperature independent, but have a correction term proportional to $T^2$. In other words, temperature not only shifts the $G(V)$ curves but also changes their curvatures. This fact has not been pointed out before. 

One more note concerns the relationship between the voltage and temperature coefficients $V_0$ and $T_0$: Their ratio is a constant with still the same value as with the bare Simmons coefficients\cite{Gloos2000}
\begin{equation}
\frac{V_0^2}{T_0^2} = \frac{V_{0,0}^2}{T_{0,0}^2} = \frac{4\pi^2}{3}\frac{k_B^2}{e^2}.\nonumber 
\label{ratio}
\end{equation}

\section{Conductance-voltage formula for rectangular barrier in the WKB approximation}

In this section, we derive an equation for conductance as a function of the bias voltage without making Simmons' fundamental approximation for the tunneling probability $D(E_x)$, Eq. \eqref{D(Ex)}, starting directly from the WKB result of Eq. \eqref{D(Ex)SecondForm}. The point of this is to have a result to which the various more approximative formulas can be compared to.  
For simplicity, let us consider a rectangular tunnel barrier whose height is $\phi_0=\phi_1=\phi_2$. A bias voltage $V$ tilts the barrier, making its shape again trapezoidal, $\phi(x)=\phi_0-eVx/d_\textrm{phys}$, where we still assume $V<\phi_0/e$.  The WKB tunneling probability, Eq. \eqref{D(Ex)SecondForm}, can then be integrated analytically (Appendix \ref{WKBGV}), giving
\begin{gather}
D_{\textrm{WKB}}(V, E_x) = \exp\Bigg\{-\frac{A'}{eV}\Big[(\mu+\phi_0-E_x)^{\frac{3}{2}} \nonumber\\ 
-(\mu+\phi_0-E_x-eV)^{\frac{3}{2}}\Big]\Bigg\}
\label{D(eV,Ex)}
\end{gather}
with $A' = 8\pi d\sqrt{2m_{\textrm{e}}}/(3h)$, where $d$ still denotes the effective thickness $d = d_{\textrm{eff}}$, see Eq. \eqref{dphys}. By substituting Eq. \eqref{D(eV,Ex)} into the general equation Eq. \eqref{WKBJ(V)Integral} and differentiating with respect to voltage (see details in Appendix \ref{WKBGV}), we get an equation for the zero-temperature conductance vs. voltage in the WKB approximation:
\begin{gather}
\frac{G}{A_\textrm{j}} = Be\Bigg(\int_{0}^{\mu-eV}D_{\textrm{WKB}}(V,E_x)\Bigg\{1+\frac{A'}{eV}\Bigg[(\mu+\phi_0-E_x)^{\frac{3}{2}}\nonumber\\
-(\mu+\phi_0-E_x-eV)^{\frac{3}{2}}-\frac{3eV}{2}(\mu+\phi_0-E_x-eV)^{\frac{1}{2}}\Bigg]\Bigg\}dE_x\nonumber\\
+\int_{\mu-eV}^{\mu}(\mu-E_x)D_{\textrm{WKB}}(V,E_x)\frac{A'}{(eV)^2}\Bigg[(\mu+\phi_0-E_x)^{\frac{3}{2}}\nonumber\\
-(\mu+\phi_0-E_x-eV)^{\frac{3}{2}}-\frac{3eV}{2}(\mu+\phi_0-E_x-eV)^{\frac{1}{2}}\Bigg]dE_x\Bigg),
\label{WKB G(V,0)}
\end{gather}
where the energy integrals have to be computed numerically. 


\section{Comparison of conductance-voltage formulas and experimental fitting}

\begin{figure}[]
\includegraphics[width=1.0\linewidth]{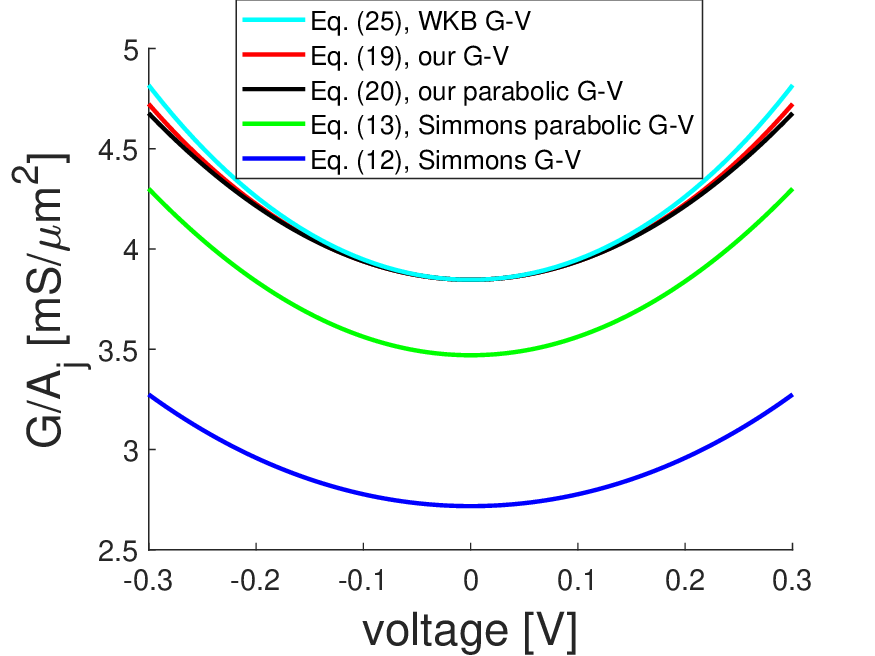}
\caption{\label{ComparisonOfGVModels5CurvesShorterVRange} Comparison of different $G-V$ models for a rectangular barrier with realistic barrier parameters $d = 9 \textrm{ Å}$ and $\phi_0=1 \textrm{ eV}$. The cyan curve shows the WKB numerical result Eq. \eqref{WKB G(V,0)} with $\mu=11.7$ eV (the Fermi energy of Al), the red curve corresponds to our Eq. \eqref{ImprovedG(V,T)} at $T=0$, the black curve shows its quadratic approximation Eq. \eqref{ParabolicApproximationOfImprovedG(V,T)} at $T=0$, whereas the green curve illustrates the parabolic Simmons model of Eq. \eqref{SimmonsSimpleGVatT=0}, and the blue curve shows the full Simmons model, Eq. \eqref{SimmonsGV}.}
\end{figure}

To demonstrate the usefulness of the improved analytical formulas of this study, in Fig.~\ref{ComparisonOfGVModels5CurvesShorterVRange} we compare the $G(V)$ curves obtained from Eqs. \eqref{ImprovedG(V,T)} and \eqref{ParabolicApproximationOfImprovedG(V,T)} in the limit $T=0$ with the direct numerical WKB results using Eq. \eqref{WKB G(V,0)}, for the rectangular barrier model. In addition, we also plot in Fig.~\ref{ComparisonOfGVModels5CurvesShorterVRange} the Simmons model results obtained from Eq. \eqref{SimmonsGV} and the parabolic approximation, Eq. \eqref{SimmonsSimpleGVatT=0}. The barrier parameters chosen, $d=9\textrm{ Å}$ and $\phi_0=1 \textrm{ eV}$, represent a typical AlOx tunneling barrier, and the voltage range $-0.3\textrm{ V }...+0.3\textrm{ V}$ roughly corresponds to a range where the parabolic approximation, Eq. \eqref{ParabolicApproximationOfImprovedG(V,T)}, starts to slightly deviate from the result of Eq. \eqref{ImprovedG(V,T)}.  

The main observation from Fig.~\ref{ComparisonOfGVModels5CurvesShorterVRange} is that our own $G-V$ formulas, Eqs. \eqref{ImprovedG(V,T)} and \eqref{ParabolicApproximationOfImprovedG(V,T)}, approximate the direct numerical WKB result well in this voltage range, and that the simplest parabolic result, Eq. \eqref{ParabolicApproximationOfImprovedG(V,T)} is already a very good practical approximation. In contrast, the curves of the Simmons model, Eqs. \eqref{SimmonsGV} and \eqref{SimmonsSimpleGVatT=0}, are in this case quite poor approximations of the WKB result. 
Eq. \eqref{SimmonsGV} gives a value more than 25 \% below Eq. \eqref{WKB G(V,0)} at $V=0$, Eq. \eqref{SimmonsSimpleGVatT=0} gives a result that is somewhat closer to Eq. \eqref{WKB G(V,0)} than Eq. \eqref{SimmonsGV}, but only by accident. Mathematically, Eq. \eqref{SimmonsSimpleGVatT=0} is not a direct Taylor expansion of Eq. \eqref{SimmonsGV} because some terms in the expansion are neglected, as discussed in detail in Appendix \ref{SimmonsJVGV}.   

\subsection{Relevance to fitting experimental data}
\begin{figure}[]
\includegraphics[width=1.0\linewidth]{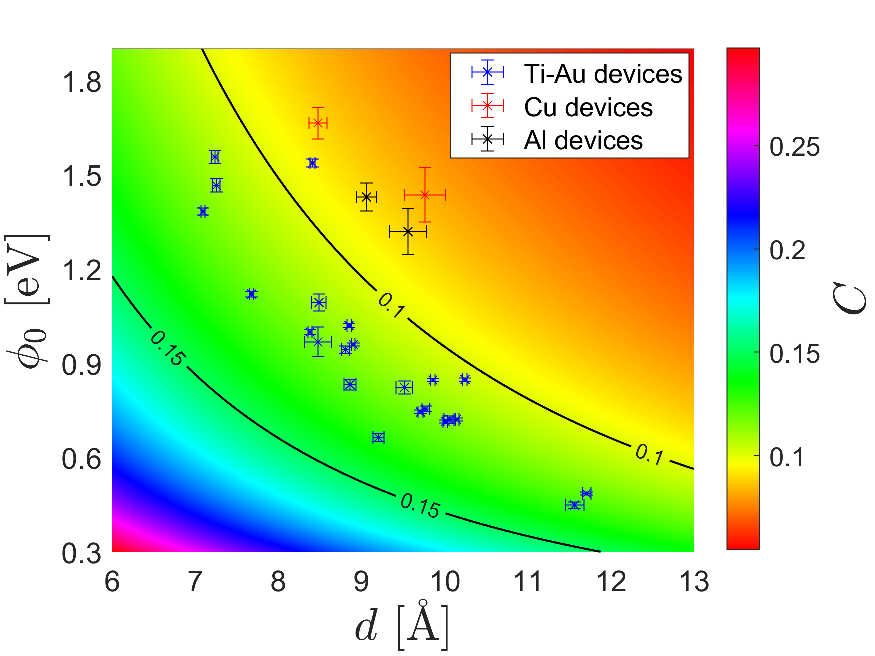}
\caption{\label{GVFitParametersRTZBWithCAndNewTheory} The correction factor $C$ as a function of $d$ and $\phi_0$ (color scale) together with fit parameters  obtained for Ti-Au (blue points), Cu (red points) and Al devices (black points) by fitting Eq. \eqref{ParabolicApproximationOfImprovedG(V,T)} to experimental $G-V$ data taken at room temperature.}
\end{figure}

Finally, we would like to address the important question of how big an impact the new Equation \eqref{ParabolicApproximationOfImprovedG(V,T)} has on fitting real experimental tunnel junction conductance data. This is first studied theoretically in Fig. \ref{GVFitParametersRTZBWithCAndNewTheory} by plotting the value of the dimensionless correction factor $C$, Eq. \eqref{C}, (color scale) as a function of both the effective barrier thickness $d$ and the average zero-bias barrier height $\phi_0$ for a realistic range of parameters. We see that the size of the correction increases for thin and shallow barriers, being fairly significant in size in the range 10 - 15 \%.     

To make a connection to experimental data, we have fitted experimental $G-V$ data measured on three different types of metal-insulator-metal tunnel junctions with AlOx barriers to the new model of Eq. \eqref{ParabolicApproximationOfImprovedG(V,T)}, with $d$ and $\phi_0$ as the two fitting parameters. All junctions had a base electrode of Al, which was oxidized in pure oxygen at room temperature, before the deposition of the counter-electrode, which was either (a) a Ti-Au bilayer, (b) Cu, or (c) Al. Details of the samples, their fabrication using electron-beam lithography and the measurements are presented elsewhere\cite{Ilmoexperimental}. 

\begin{figure}[]
\includegraphics[width=1.0\linewidth]{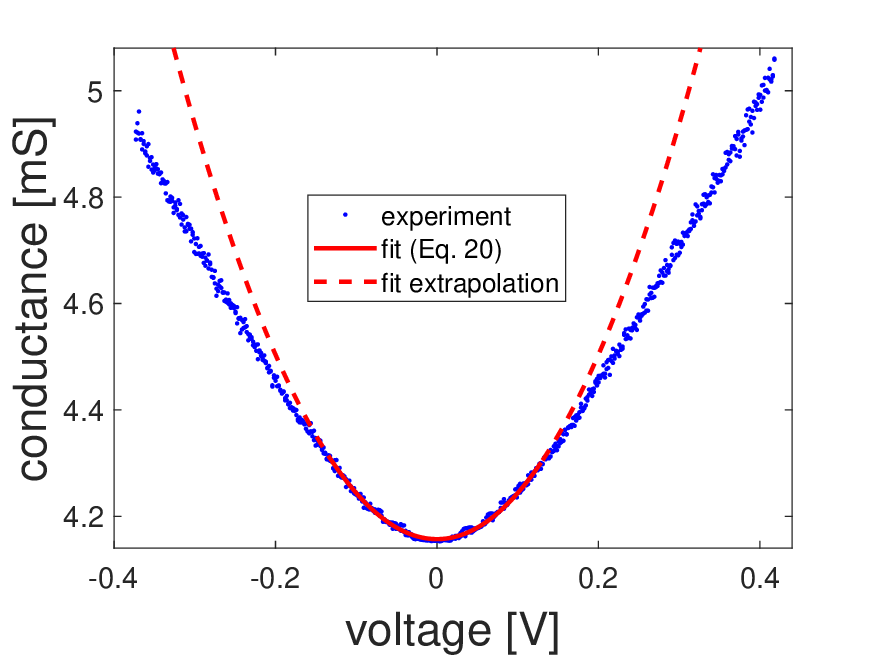}
\caption{\label{GVLargeTiAuRTOneJunction} Experimental $G-V$ characteristic (blue dots) of a representative Ti-Au device with $A_{\text{j}}=(0.52\pm0.01)\text{ }\mu\text{m}^2$ at room temperature. The solid red line at $|V| \leq 0.13 \textrm{ V}$ shows the fit given by Eq. \eqref{ParabolicApproximationOfImprovedG(V,T)}. From the fit, one can calculate barrier parameters $d=(8.38\pm0.03)\text{ Å}$ and $\phi_0=(1.00\pm0.01)\text{ eV}$. The dashed red line at $|V| > 0.13 \textrm{ V}$ shows the extrapolation of the fit.}
\end{figure}

Fig. \ref{GVLargeTiAuRTOneJunction} shows an example of room temperature experimental $G-V$ data from a Ti-Au device with a fit to the model of Eq. \eqref{ParabolicApproximationOfImprovedG(V,T)} in the voltage range $|V| \leq 0.13 \textrm{ V}$. The junction area $A_{\text{j}}$ was not used as a fit parameter but was always determined separately using scanning electron microscopy of the devices.
Appendix \ref{CalculationOfBarrierParameters} explains the details of the fitting procedure and the analysis of parameter errors.

Looking back at Fig. \ref{GVFitParametersRTZBWithCAndNewTheory}, it also shows a collection of the fitted parameters $d$ and $\phi_0$ for all three device types. For Cu and Al devices, the obtained barrier heights $\phi_0 > 1.2$ eV are typically larger than for Ti-Au devices, for which in most cases $\phi_0 \approx 0.7 - 1.1 $ eV. The correction factor $C$ for Cu and Al devices is somewhat below $C=0.1$, whereas for almost all Ti-Au devices, $C>0.1$. Such values indicate a significant difference between the standard Simmons model and our improved version for real devices. In addition to affecting the parameter fit values, their errors are also decreased by using the new model: The errors of $d$ are $0\text{ }\%$ - $30\text{ }\%$ smaller and the errors of $\phi_0$ are $50\text{ }\%$ - $70\text{ }\%$ smaller when the fitting is done with the new Eq. \eqref{ParabolicApproximationOfImprovedG(V,T)} instead of the Simmons model Eq. \eqref{SimmonsSimpleGVatT=0}.

\section{Conclusions}
In conclusion, we have derived a new analytical formula for the current density through a tunnel barrier of arbitrary shape at finite voltage and temperature, Eq. \eqref{ImprovedGeneralizedJ(V,T)}, and formulas for the conductance of a trapezoidal barrier for bias voltages smaller than the lower of the two barrier heights $0\leq V \leq \phi_1/e$, including simultaneously the effect of a finite temperature, Eqs. \eqref{ImprovedG(V,T)} and \eqref{ParabolicApproximationOfImprovedG(V,T)}.
Eq. \eqref{ParabolicApproximationOfImprovedG(V,T)} shows that the effect of finite temperature is twofold: It not only shifts the zero-bias value of the conductance, a previously known effect, but it also changes the curvature of the conductance-voltage curve, which has not been understood before. 

We also demonstrated that the new conductance-voltage formulas are much more accurate approximations of the numerically evaluated Wentzel-Kramers-Brillouin approximation than the well-known Simmons model of Eq. \eqref{SimmonsSimpleGVatT=0}, for realistic barrier parameters. 

Such simple formulas can be used to fit experimental data. In particular, we demonstrated how to use the quadratic formula of Eq. \eqref{ParabolicApproximationOfImprovedG(V,T)} in linear least-squares fitting of experimental conductance-voltage data from several different types of metal-insulator-metal tunnel junctions. For the junctions studied, the new more accurate formula predicts a correction of approximately $\sim 10$ \% to the key parameters of the old Simmons model, showing that the improved formulas have practical relevance when determining the barrier properties of tunnel junctions.

\begin{acknowledgments}

We thank Z. Geng and K. M. Kinnunen for technical assistance in experiments. 
This work has been funded by Finnish Cultural Foundation, the Vilho, Yrjö and Kalle Väisälä Foundation, and the Research Council of Finland project number 341823.
\end{acknowledgments}

\section*{Author declarations}

\subsection*{Conflict of Interest}
The authors have no conflicts to disclose.
\subsection*{Author Contributions}
\textbf{I. M. W. Räisänen:} Data curation (lead); Formal analysis (equal); Funding acquisition (equal); Investigation (equal); Software (lead); Visualization (lead); Writing \textendash\text{ }original draft (lead); Writing \textendash\text{ }review \& editing (equal).
\textbf{I. J. Maasilta:} Conceptualization (lead); Formal analysis (equal); Funding acquisition (equal); Investigation (equal); Methodology (lead); Project administration (lead); Supervision (lead); Writing \textendash\text{ }review \& editing (equal).

\section*{Data Availability Statement}

The data that support the findings of this study are available from the corresponding author upon reasonable request.

\appendix

\section{\label{SimmonsJVGV} Re-derivation of Simmons' zero-temperature $J-V$ and $G-V$ equations with comments}

At zero temperature, the elastic tunneling current density $J$ as a function of voltage $V$ is described by Eq. \eqref{WKBJ(V)Integral}, i.e., 
\begin{gather}
J = B\Big[eV\int_{0}^{\mu-eV} D(E_x)dE_x
+\int_{\mu-eV}^{\mu}(\mu-E_x)D(E_x)dE_x\Big]
\label{WKBJ(V)IntegralApp}
\end{gather}
with
\begin{equation}
B=\frac{4\pi m_{\textrm{e}}e}{h^3}. \nonumber
\label{B'App}
\end{equation}
Simmons starts from the WKB approximation expression for the tunneling probability $D(E_x)$, Eq. \eqref{D(Ex)SecondForm}
\begin{equation}
D(E_x) = \exp\left[-\frac{4\pi\sqrt{2m{_\textrm{e}}}}{h}\int_{x_1}^{x_2}(\mu+\phi(x)-E_x)^{\frac{1}{2}}dx \right],
\label{D(Ex)SecondFormApp}
\end{equation}
with the barrier effective mass set to the free-electron mass $m^* = m{_\textrm{e}}$. By denoting $f(x) = \mu+\phi(x)-E_x$ and defining its average as 
\begin{equation*}
\overline{f} = \frac{1}{\Delta s}\int_{s_1}^{s_2}\!\!\!f(x)dx = \mu+\overline{\phi}-E_x,
\end{equation*}
where $\Delta s = s_2 -s_1$ and 
\begin{equation*}
\overline{\phi} = \frac{1}{\Delta s}\int_{s_1}^{s_2}\!\!\!\phi(x)dx
\end{equation*}
is the average barrier height, the integral in the WKB expression \eqref{D(Ex)SecondFormApp} can be written as 
\begin{gather}
\int_{s_1}^{s_2} f^{\frac{1}{2}}(x)dx\nonumber\\ = \int_{s_1}^{s_2}(\overline{f}+f(x)-\overline{f})^{\frac{1}{2}}dx = \overline{f}^{\frac{1}{2}}\int_{s_1}^{s_2}\left(1+\frac{f(x)-\overline{f}}{\overline{f}}\right)^{\frac{1}{2}}dx\nonumber\\
\approx\overline{f}^{\frac{1}{2}}\int_{s_1}^{s_2}\left[1+\frac{f(x)-\overline{f}}{2\overline{f}}-\frac{(f(x)-\overline{f})^2}{8\overline{f}^2}+...\right] dx \nonumber\\
= \overline{f}^{\frac{1}{2}}\Delta s\left[1-\frac{1}{8 \overline{f}^2\Delta s}\int_{s_1}^{s_2}(f(x)-\overline{f})^2dx\right]\nonumber\\
= (\mu+\overline{\phi}-E_x)^{\frac{1}{2}}\Delta s\left[1-\frac{1}{8 (\mu+\overline{\phi}-E_x)^2\Delta s}\int_{s_1}^{s_2}(\phi(x)-\overline{\phi})^2dx\right],
\label{integral}
\end{gather} 
where the third line is a Taylor expansion assuming $(f(x)-\overline{f})/\overline{f} \ll 1$, and the fourth line follows after the integration of the constant and quadratic terms, with the linear term giving a zero. By defining the correction factor in the brackets as 
\begin{equation*}
\beta = 1-\frac{1}{8 (\mu+\overline{\phi}-E_x)^2\Delta s}\int_{s_1}^{s_2}(\phi(x)-\overline{\phi})^2dx,
\end{equation*}
and substituting Simmons' approximation for the WKB integral, Eq. \eqref{integral}, into the expression for the WKB tunneling probability, Eq. \eqref{D(Ex)SecondFormApp}, we arrive at  
Simmons' approximation for the tunneling probability, Eq. \eqref{D(Ex)} in the main text, i.e.,
\begin{equation}
D(E_x) \simeq \exp\left[-A(\mu+\overline{\phi}-E_x)^{\frac{1}{2}}\right]
\label{D(Ex)App}
\end{equation}
with
\begin{equation}
A = \frac{4\pi\beta \Delta s\sqrt{2m_{\textrm{e}}}}{h}. \nonumber
\label{AApp}
\end{equation}

Then, by substituting Eq. \eqref{D(Ex)App} into Eq. \eqref{WKBJ(V)IntegralApp}, 
we get
\begin{gather}
J = B\Bigg\{eV\int_{0}^{\mu-eV} \exp[-A(\mu+\overline{\phi}-E_x)^{\frac{1}{2}}]dE_x\nonumber\\
+\int_{\mu-eV}^{\mu}(\mu-E_x)\exp[-A(\mu+\overline{\phi}-E_x)^{\frac{1}{2}}]dE_x\Bigg\}.
\label{SimilarSimmonsEq12App}
\end{gather}

To make the integration easier, Simmons writes Eq. \eqref{SimilarSimmonsEq12App} in the form
\begin{gather}
J = B\Bigg\{eV\int_{0}^{\mu-eV} \exp[-A(\mu+\overline{\phi}-E_x)^{\frac{1}{2}}]dE_x\nonumber\\
-\overline{\phi}\int_{\mu-eV}^{\mu}\exp[-A(\mu+\overline{\phi}-E_x)^{\frac{1}{2}}]dE_x\nonumber\\
+\int_{\mu-eV}^{\mu}(\mu+\overline{\phi}-E_x)\exp[-A(\mu+\overline{\phi}-E_x)^{\frac{1}{2}}]dE_x\Bigg\}
\label{SimilarSimmonsEq13App}
\end{gather}
which is Eq. (13) of Ref. \onlinecite{SimmonsSimilarElectrodes}. Assuming $\beta$ to be constant, the first term of Eq. \eqref{SimilarSimmonsEq13App} integrates to Eq. (14) of Ref. \onlinecite{SimmonsSimilarElectrodes} which is
\begin{gather}
(2BeV/A^2)\{[A({\overline{\phi}}+eV)^{\frac{1}{2}}+1]\exp[-A({\overline{\phi}}+eV)^{\frac{1}{2}}]\nonumber\\
-[A(\overline{\phi}+\mu)^{\frac{1}{2}}+1]\exp[-A(\overline{\phi}+\mu)^{\frac{1}{2}}]\}.
\label{SimilarSimmonsEq14App}
\end{gather}

Assuming good metals, for which typically $\mu > eV$, the second term in the braces in Eq. \eqref{SimilarSimmonsEq14App} is small compared to the first term and can be neglected. The second term of Eq. \eqref{SimilarSimmonsEq13App} is equal to Eq. (16) of Ref. \onlinecite{SimmonsSimilarElectrodes} which is
\begin{gather}
(2B\overline{\phi}/A^2)\{[A({\overline{\phi}}+eV)^{\frac{1}{2}}+1]\exp[-A({\overline{\phi}}+eV)^{\frac{1}{2}}]\nonumber\\
-[A\overline{\phi}^{\frac{1}{2}}+1]\exp[-A\overline{\phi}^{\frac{1}{2}}]\}.
\label{SimilarSimmonsEq16App}
\end{gather}

By defining $z=(\mu+\overline{\phi}-E_x)^{\frac{1}{2}}$, the third term of Eq. \eqref{SimilarSimmonsEq13App} becomes
\begin{eqnarray}
&-& 2B\int_{(\overline{\phi}+eV)^{\frac{1}{2}}}^{\overline{\phi}^{\frac{1}{2}}}z^3\exp(-Az)dz \nonumber\\
&=& 2B\exp(-Az)\Big(\frac{z^3}{A}+\frac{3z^2}{A^2}+\frac{6z}{A^3}+\frac{6}{A^4}\Big) \bigg|_{(\overline{\phi}+eV)^{\frac{1}{2}}}^{\overline{\phi}^{\frac{1}{2}}}.
\label{SimmonsJ(V)IntegralThirdIntegralTrick}
\end{eqnarray}

Simmons simply neglects the terms $6z/A^3$ and $6/A^4$ in Eq. (17) of Ref. \onlinecite{SimmonsSimilarElectrodes}, or Eq. \eqref{SimmonsJ(V)IntegralThirdIntegralTrick} of this paper. Eq. \eqref{SimmonsJ(V)IntegralThirdIntegralTrick} then becomes Eq. (18) of Ref. \onlinecite{SimmonsSimilarElectrodes} which is
\begin{gather}
(2B/A)\{\overline{\phi}^{\frac{3}{2}}\exp(-A\overline{\phi}^{\frac{1}{2}})-(\overline{\phi}+eV)^{\frac{3}{2}}\exp[-A(\overline{\phi}+eV)^{\frac{1}{2}}]\} \nonumber\\
+(6B/A^2)\{\overline{\phi}\exp(-A\overline{\phi}^{\frac{1}{2}}) \nonumber\\
-(\overline{\phi}+eV)\exp[-A(\overline{\phi}+eV)^{\frac{1}{2}}]\}. 
\label{SimilarSimmonsEq18App}
\end{gather}

In Appendix \ref{J(V,0)} we show that neglecting terms $6z/A^3$ and $6/A^4$ is the approximation that makes Simmons' zero-temperature equations, i.e., Eqs. \eqref{SimmonsGeneralizedJV} and \eqref{SimmonsGV}, different from our Eqs. \eqref{ImprovedGeneralizedJ(V,T)} and \eqref{ImprovedG(V,T)} at $T=0$.

Summing Eq. \eqref{SimilarSimmonsEq14App} without its second term $-[A(\overline{\phi}+\mu)^{\frac{1}{2}}+1]\exp[-A(\overline{\phi}+\mu)^{\frac{1}{2}}]$, Eq. \eqref{SimilarSimmonsEq16App} and Eq. \eqref{SimilarSimmonsEq18App}, Simmons gets Eq. (20) of Ref. \onlinecite{SimmonsSimilarElectrodes} which is
\begin{equation}
J = J_0\{\overline{\phi}\exp(-A{\overline{\phi}}^{\frac{1}{2}})
-(\overline{\phi}+eV)\exp[-A({\overline{\phi}}+eV)^{\frac{1}{2}}]\}
\label{SimmonsGeneralizedJVOriginalForm}
\end{equation}
with
\begin{equation}
J_0 = \frac{e}{2\pi h(\beta \Delta s)^2}. \nonumber
\label{J_0}
\end{equation}
Since $J_0=4B/A^2$, Eq. \eqref{SimmonsGeneralizedJVOriginalForm} is equivalent to Eq. \eqref{SimmonsGeneralizedJV}. Unlike what is written in Ref. \onlinecite{SimmonsSimilarElectrodes}, the approximation $A(\overline{\phi}+eV)^{\frac{1}{2}}\gg1$ leading to Eq. (15) of Ref. \onlinecite{SimmonsSimilarElectrodes} is not necessary in the derivation of Eq. \eqref{SimmonsGeneralizedJVOriginalForm}. 

Considering trapezoidal barriers in the low-voltage range $eV < \phi_0$, i.e. by making the approximation $\beta=1$ and substituting Eq. \eqref{overline_phi2}, i.e.,
\begin{equation}
\left\{ \begin{array}{ll}
\overline{\phi} = \phi_0 - \frac{eV}{2}\\
\Delta s = d \nonumber
\label{overline_phi2App}
\end{array}\right.
\end{equation}
into Eq. \eqref{SimmonsGeneralizedJVOriginalForm}, Simmons gets Eq. (27) of Ref. \onlinecite{SimmonsSimilarElectrodes} which is
\begin{eqnarray}
J &=& \frac{e}{2\pi hd^{2}}\Bigg\{\left(\phi_0-\frac{eV}{2}\right)\exp\left[-A\left(\phi_0-\frac{eV}{2}\right)^{\frac{1}{2}}\right]\nonumber\\
&\text{ }& - \left(\phi_0+\frac{eV}{2}\right)\exp\left[-A\left(\phi_0+\frac{eV}{2}\right)^{\frac{1}{2}}\right]\Bigg\},
\label{SimmonsJV}
\end{eqnarray}
with $A=2d\sqrt{2m_{\text{e}}}/\hbar$. Differentiating Eq. \eqref{SimmonsJV} with respect to voltage, one gets Eq. \eqref{SimmonsGV} which is
\begin{gather}
\frac{G}{A_\textrm{j}} = \frac{e^2}{8\pi hd^{2}}\Bigg\{\exp\left[-A\left(\phi_0-\frac{eV}{2}\right)^{\frac{1}{2}}\right]\left[A\left(\phi_0-\frac{eV}{2}\right)^{\frac{1}{2}}-2\right]\nonumber\\
+ \exp\left[-A\left(\phi_0+\frac{eV}{2}\right)^{\frac{1}{2}}\right]\left[A\left(\phi_0+\frac{eV}{2}\right)^{\frac{1}{2}}-2\right]\Bigg\}.
\label{SimmonsGVappx}
\end{gather}
The second-degree Taylor polynomial of Eq. \eqref{SimmonsGVappx} around $V=0$ is
\begin{eqnarray}
\frac{G}{A_\textrm{j}} &\approx& \frac{e^2\exp(-A\phi_0^{\frac{1}{2}})}{4\pi hd^2}(A\phi_0^{\frac{1}{2}}-2) \nonumber \\
&\text{ }& + \frac{Ae^4\exp(-A\phi_0^{\frac{1}{2}})}{128\pi hd^2\phi_0^{\frac{3}{2}}}(A^2\phi_0-3A\phi_0^{\frac{1}{2}}-3)V^2.
\label{SimmonsGVTaylorSeriesMoreDetails}
\end{eqnarray}
To simplify this result, one could neglect the smaller terms $-2$, $-3A\phi_0^{1/2}$ and $-3$ in the coefficient polynomials of Eq. \eqref{SimmonsGVTaylorSeriesMoreDetails}, to get the lowest-order result
\begin{eqnarray}
\frac{G}{A_\textrm{j}} &\approx& \frac{e^2\exp(-A\phi_0^{\frac{1}{2}})}{4\pi hd^2}A\phi_0^{\frac{1}{2}}+\frac{Ae^4\exp(-A\phi_0^{\frac{1}{2}})}{128\pi hd^2\phi_0^{\frac{3}{2}}}A^2\phi_0V^2 \nonumber \\
&=& \frac{e^2A\phi_0^{\frac{1}{2}}}{4\pi hd^2}\exp(-A\phi_0^{\frac{1}{2}})\left(1+\frac{e^2A^2}{32\phi_0}V^2\right) \nonumber \\
&=& \frac{e^2\sqrt{2m_{\textrm{e}}\phi_0}}{h^2d}\exp\left[-\frac{2d\sqrt{2m_{\textrm{e}}\phi_0}}{\hbar}\right] \nonumber \\
&\text{ }& \times \left(1+\frac{e^2m_{\textrm{e}}d^2}{4\hbar^2\phi_0}V^2\right), \nonumber
\label{SimmonsGVTaylorSeriesSimplification}
\end{eqnarray}
which is equivalent to Eq. \eqref{SimmonsSimpleGVatT=0}, i.e.,
\begin{equation}
G(V,T=0) = G_{0,0}\left(1+\frac{V^2}{V_{0,0}^2}\right)
\label{SimmonsSimpleGVatT=0App}
\end{equation}
with
\begin{equation} 
G_{0,0} = \frac{e^2A_\textrm{j}\sqrt{2m_\textrm{e}\phi_0}}{h^2d}\textrm{exp}\biggr[\frac{-2d\sqrt{2m_\textrm{e}\phi_0}}{\hbar}\biggr] \nonumber
\label{G_00App}
\end{equation}
and
\begin{equation}
V_{0,0}^2 = \frac{4\hbar^2\phi_0}{e^2m_\textrm{e}d^2}. \nonumber
\label{V_00App}
\end{equation}

However, neglect of the terms $-2$, $-3A\phi_0^{1/2}$ and $-3$ in Eq. \eqref{SimmonsGVTaylorSeriesMoreDetails} leading to Eq. \eqref{SimmonsSimpleGVatT=0App} cannot always be well justified. For barrier parameters $d=15\text{ Å}$ and $\phi_0=2\textrm{ eV}$ that Simmons uses in Ref. \onlinecite{SimmonsLowVoltage}, one gets $A\phi_0^{1/2} \approx 22$ and $A^2\phi_0 \approx 470$. For parameters $d=9\text{ Å}$ and $\phi_0=1\textrm{ eV}$ that are usual for Ti-Au devices according to Fig. \ref{GVFitParametersRTZBWithCAndNewTheory}, we get $A\phi_0^{1/2} \approx 9.2$ and $A^2\phi_0 \approx 85$, leading to an even bigger error if using Eq. \eqref{SimmonsSimpleGVatT=0} (Eq. \eqref{SimmonsSimpleGVatT=0App}). However, we stress that keeping the above terms will not improve the accuracy (as demonstrated in Fig.~\ref{ComparisonOfGVModels5CurvesShorterVRange} in the main text), as Eq. \eqref{SimmonsGVappx} already contains approximations causing errors of the same order of magnitude.   

To comment on some other explicit equations in Simmons' papers, we can look at the third-degree Taylor polynomial of Eq. \eqref{SimmonsJV} around $V=0$, which is
\begin{eqnarray}
J &\approx& \frac{e^2\exp(-A\phi_0^{\frac{1}{2}})}{4\pi hd^2}(A\phi_0^{\frac{1}{2}}-2)V \nonumber \\
&\text{ }& + \frac{Ae^4\exp(-A\phi_0^{\frac{1}{2}})}{384\pi hd^2\phi_0^{\frac{3}{2}}}(A^2\phi_0-3A\phi_0^{\frac{1}{2}}-3)V^3.
\label{SimmonsJVTaylorSeriesMoreDetails}
\end{eqnarray}
It is almost identical to Eq. (3) of Ref. \onlinecite{SimmonsLowVoltage}, but in Eq. (3) of Ref. \onlinecite{SimmonsLowVoltage} there is an incorrect prefactor of $3/2$ that should be $2$. If one neglects the terms $-2$ and $-3$ but takes into account the terms $A^2\phi_0$ and $-3A\phi_0^{1/2}$ in Eq. \eqref{SimmonsJVTaylorSeriesMoreDetails}, one gets
\begin{eqnarray}
J &\approx& \frac{e^2\exp(-A\phi_0^{\frac{1}{2}})}{4\pi hd^2}A\phi_0^{\frac{1}{2}}V+\frac{Ae^4\exp(-A\phi_0^{\frac{1}{2}})}{384\pi hd^2\phi_0^{\frac{3}{2}}}(A^2\phi_0-3A\phi_0^{\frac{1}{2}})V^3 \nonumber \\
&=& \frac{e^2A\phi_0^{\frac{1}{2}}}{4\pi hd^2}\exp(-A\phi_0^{\frac{1}{2}})\left[V+\frac{e^2}{96\phi_0^2}(A^2\phi_0-3A\phi_0^{\frac{1}{2}})V^3\right] \nonumber \\
&=& \frac{e^2\sqrt{2m_{\textrm{e}}\phi_0}}{h^2d}\exp(-A\phi_0^{\frac{1}{2}})\Bigg\{V+\Bigg[\frac{(Ae)^2}{96\phi_0}-\frac{Ae^2}{32\phi_0^{\frac{3}{2}}}\Bigg]V^3\Bigg\},
\label{SimmonsJVTaylorSeriesSimplification}
\end{eqnarray}
which is equivalent to Eq. (13a) of Ref. \onlinecite{SimmonsDissimilarElectrodes} with the assumption $\phi_0=(\phi_1+\phi_2)/2$. Eq. (4) of Ref. \onlinecite{SimmonsLowVoltage} is almost identical to Eq. \eqref{SimmonsJVTaylorSeriesSimplification}, but in Eq. (4) of Ref. \onlinecite{SimmonsLowVoltage} there is an incorrect prefactor of $3/2$ which should be $1$. As mentioned in the main text, Eqs. \eqref{SimmonsGVappx} and \eqref{SimmonsSimpleGVatT=0App} are not shown in Simmons' papers \cite{SimmonsLowVoltage, SimmonsSimilarElectrodes, SimmonsDissimilarElectrodes, SimmonsThermal}.

We also report the following inconsistencies and errors in Simmons' equations. These inconsistencies can be verified by performing simple algebra. In our statements \textbf{1)}-\textbf{4)}, we refer to the equations of Ref. \onlinecite{SimmonsLowVoltage} unless we write something else:\\
\\
\textbf{1)} The latter term of Eq. (2), the starting point of Ref. \onlinecite{SimmonsLowVoltage}, is not equivalent to Eq. (27) of Ref. \onlinecite{SimmonsSimilarElectrodes} (Eq. \eqref{SimmonsJV} of this article). They would be equivalent if the prefactor of the latter term of Eq. (2) were $2$ instead of $3$.\\
\\
\textbf{2)} Eq. (3) is not consistent with its preceding equations. It would be consistent if the prefactor in Eq. (3) were $3$ instead of $3/2$.\\ 
\\
\textbf{3)} Eq. (4) is not consistent with Eq. (3). They would be consistent if the prefactor in the coefficient $\beta$ were $3/4$ instead of $3/2$. To be consistent with the correct Eq. (27) of Ref. \onlinecite{SimmonsSimilarElectrodes}, the prefactor should be $1$, as already discussed after Eq. \eqref{SimmonsJVTaylorSeriesSimplification} of this article.\\
\\
\textbf{4)} Eq. (5), which can be obtained by neglecting the cubic term in Eq. (4), is identical to Eq. (25) of Ref. \onlinecite{SimmonsSimilarElectrodes}, but they are incorrect\cite{SimmonsCorrection}. The prefactor $3/2$ in Eq. (25) of Ref. \onlinecite{SimmonsSimilarElectrodes} and in the coefficient $\beta$ in Ref. \onlinecite{SimmonsLowVoltage} should be $1$. \\
\\
Some results of Simmons' papers are also reviewed in Wolf's book, Ref. \onlinecite{Wolf}, on which we report the following errors, which can be verified by doing simple algebra. In our statements \textbf{5)}-\textbf{6)}, we refer to the equations of Ref. \onlinecite{Wolf} unless we write something else:\\
\\
\textbf{5)} Eq. (2.47a) together with Eq. (2.51) agrees with Eq. \eqref{SimmonsJVTaylorSeriesSimplification} of this article only if the exponent of $\overline{\phi}$ in the latter term of the quantity $\gamma/\alpha$ is $3/2$ instead of $1/2$.\\
\\
\textbf{6)} As Eq. (2.47a) describes the current density, Eq. (2.47b) describes the conductance per junction area, not the conductance itself.\\

\section{\label{SimmonsJTGT} Re-derivation of Simmons' finite-temperature $J-T$ and $G-T$ equations with comments}

At nonzero temperatures, the elastic tunneling current density $J$ as a function of voltage $V$ and temperature $T$ is described by Eq. \eqref{WKBJ(V,T)Integral}, i.e.,
\begin{gather}
J(V,T) = Bk_{\textrm{B}}T\int_{0}^{E_m} D(E_x)\nonumber\\
\times \ln \Bigg\{\frac{1+\exp[(\mu-E_x)/(k_{\textrm{B}}T)]}{1+\exp[(\mu-E_x-eV)/(k_{\textrm{B}}T)]}\Bigg\}dE_x
\label{WKBJ(V,T)IntegralAppB}
\end{gather}
with
\begin{equation}
B=\frac{4\pi m_{\textrm{e}}e}{h^3}. \nonumber
\label{B'AppB}
\end{equation}
Simmons' approximation for the tunneling probability, Eq. \eqref{D(Ex)} (derived in Appendix \ref{SimmonsJVGV}), can be written in the form
\begin{eqnarray}
D(E_x) &\simeq& \exp\left[-A(\mu+\overline{\phi}-E_x)^{\frac{1}{2}}\right] \nonumber \\
&=& \exp\left[-A\overline{\phi}^{\frac{1}{2}}\left(1+\frac{\mu-E_x}{\overline{\phi}}\right)^{\frac{1}{2}}\right],
\label{D(Ex)AppB}
\end{eqnarray}
with
\begin{equation}
A = \frac{4\pi\beta \Delta s\sqrt{2m_{\textrm{e}}}}{h}. \nonumber
\label{AAppB}
\end{equation}

To be able to integrate Eq. \eqref{WKBJ(V,T)IntegralAppB} analytically, Simmons makes an additional approximation compared to his zero-temperature formulas, by assuming $(\mu-E_x)/\overline{\phi} \ll 1$, which allows him to make the lowest-order Taylor expansion such that  
\begin{equation}
\left(1+\frac{\mu-E_x}{\overline{\phi}}\right)^{\frac{1}{2}} \approx 1+\frac{1}{2}\frac{\mu-E_x}{\overline{\phi}}.
\label{SimmonsTapproximation}
\end{equation}
This he justifies by stating in Ref. \onlinecite{SimmonsThermal} that only electrons with energies $E_x$ close to $\mu$ contribute effectively to the tunneling current. However, this statement is not mathematically accurate for realistic barrier parameters and temperatures, which can be seen by studying the integrand of Eq. \eqref{WKBJ(V,T)IntegralAppB} numerically. 

Nevertheless, by making the above rough approximation, Eq. \eqref{D(Ex)AppB} becomes Eq. (6) of Ref. \onlinecite{SimmonsThermal} which is
\begin{equation}
D(E_x) \approx \exp(-A\overline{\phi}^{\frac{1}{2}})\exp\left[-A(\mu-E_x)/(2\overline{\phi}^{\frac{1}{2}})\right].
\label{D(Ex)SimmonsApproximation}
\end{equation}
By substituting Eq. \eqref{D(Ex)SimmonsApproximation} into Eq. \eqref{WKBJ(V,T)IntegralAppB}, extending the upper limit to infinity and the lower limit to minus infinity
and integrating as in Ref. \onlinecite{Murphy}, Simmons gets Eq. (8) of Ref. \onlinecite{SimmonsThermal} which is
\begin{gather}
J(V,T) = \frac{B}{\overline{B}^2}\frac{\pi \overline{B}k_{\textrm{B}}T}{\sin (\pi \overline{B}k_{\textrm{B}}T)}\nonumber\\
\times \exp(-A\overline{\phi}^{\frac{1}{2}})[1-\exp(-\overline{B}eV)],
\label{SimmonsJ(V,T)AfterIntegration}
\end{gather}
where $\overline{B}=A/(2{\overline{\phi}}^{\frac{1}{2}})$. At $T=0$, Eq. \eqref{SimmonsJ(V,T)AfterIntegration} becomes the first line in Eq. (9) of Ref. \onlinecite{SimmonsThermal}, which is
\begin{eqnarray}
J(V,0) &=& (B/\overline{B}^2)\exp(-A\overline{\phi}^{\frac{1}{2}})[1-\exp(-\overline{B}eV)] \nonumber \\
&=& \frac{4B}{A^2}\Bigg\{\overline{\phi}\exp(-A{\overline{\phi}}^{\frac{1}{2}}) \nonumber \\
&\text{ }& - \overline{\phi}\exp\Bigg[-A\overline{\phi}^{\frac{1}{2}}\left(1+\frac{eV}{2\overline{\phi}}\right)\Bigg]\Bigg\}.
\label{SimmonsGeneralizedJVAppB}
\end{eqnarray}
Simmons further assumes that $eV\ll\overline{\phi}$, which allows him to add all higher-order Taylor polynomial terms to the linear approximation such that
\begin{equation}
1+\frac{eV}{2\overline{\phi}} \approx \left(1+\frac{eV}{\overline{\phi}}\right)^{\frac{1}{2}}, \nonumber
\label{SimmonsLowVoltageApproximation}
\end{equation}
which allows Eq. \eqref{SimmonsGeneralizedJVAppB} to be written as
\begin{equation}
J(V,0) \approx \frac{4B}{A^2}\Big\{\overline{\phi}\exp(-A{\overline{\phi}}^{\frac{1}{2}})
-\overline{\phi}\exp[-A({\overline{\phi}}+eV)^{\frac{1}{2}}]\Big\}
\label{SimmonsGeneralizedJVAppBSecondForm}
\end{equation}
corresponding to the second line of Eq. (9) of Ref. \onlinecite{SimmonsThermal}.
Eq. \eqref{SimmonsGeneralizedJVAppBSecondForm} is otherwise identical to Simmons' zero-temperature result Eq. (20) of Ref. \onlinecite{SimmonsSimilarElectrodes}, presented in Eq. \eqref{SimmonsGeneralizedJV} of this article, i.e., 
\begin{equation}
J(V,0) = \frac{4B}{A^2}\Big\{\overline{\phi}\exp(-A{\overline{\phi}}^{\frac{1}{2}})
-(\overline{\phi}+eV)\exp[-A({\overline{\phi}}+eV)^{\frac{1}{2}}]\Big\},
\label{SimmonsGeneralizedJVAppBOriginal}
\end{equation}
except that the factor multiplying the last exponential is $\overline{\phi}$ in Eq. \eqref{SimmonsGeneralizedJVAppBSecondForm} instead of $(\overline{\phi}+eV)$ in Eq. \eqref{SimmonsGeneralizedJVAppBOriginal}. Simmons himself calls this an error resulting from the approximation of Eq. \eqref{SimmonsTapproximation}, and, without mathematical rigor, simply asserts at this point that the correct $J(V,T)$ expression to be used is
\begin{equation}
J(V,T) = J(V,0)\frac{\pi \overline{B}k_{\textrm{B}}T}{\sin (\pi \overline{B}k_{\textrm{B}}T)},
\label{SimmonsJVTsin}
\end{equation}
with $J(V,0)$ given by Eq. \eqref{SimmonsGeneralizedJVAppBOriginal} instead of Eq. \eqref{SimmonsGeneralizedJVAppBSecondForm}. 

The second-order Taylor polynomial of the temperature-dependent part of Eq. \eqref{SimmonsJVTsin} around $T=0$ is
\begin{eqnarray}
\frac{\pi \overline{B}k_{\textrm{B}}T}{\sin (\pi \overline{B}k_{\textrm{B}}T)} &\approx& 1+\frac{1}{6}(\pi \overline{B}k_{\textrm{B}}T)^2 \nonumber \\
&=& 1+\frac{(\pi Ak_{\text{B}}T)^2}{24\overline{\phi}},
\label{Simmons2ndTapproximation}
\end{eqnarray}
which is a good approximation even at room temperature for typical barrier heights of the eV range. 
Using  Eq. \eqref{Simmons2ndTapproximation} in Eq. \eqref{SimmonsJVTsin} with \eqref{SimmonsGeneralizedJVAppBOriginal}, Simmons gets Eq. (11) of Ref. \onlinecite{SimmonsThermal}, which is Eq. \eqref{SimmonsGeneralizedJ(V,T)} in the main part of this article, i.e.,
\begin{equation}
J(V,T) = J(V,0)\Big[1+\frac{(\pi Ak_{\text{B}}T)^2}{24\overline{\phi}(V)}\Big],
\label{SimmonsGeneralizedJ(V,T)AppB}
\end{equation}
where $J(V,0)$ is defined by Eq. \eqref{SimmonsGeneralizedJVAppBOriginal} (Eq. \eqref{SimmonsGeneralizedJV}). 

Considering trapezoidal barriers in the voltage range $eV < \phi_0$, i.e. by making the approximation $\beta=1$ and substituting Eq. \eqref{overline_phi2}, i.e.,
\begin{equation}
\left\{ \begin{array}{ll}
\overline{\phi} = \phi_0 - \frac{eV}{2}\\
\Delta s = d
\label{overline_phi2AppB}
\end{array}\right.
\end{equation}
into Eq. \eqref{SimmonsGeneralizedJ(V,T)AppB} and differentiating with respect to voltage, one gets the specific conductance
\begin{eqnarray}
\frac{G(V,T)}{A_{\textrm{j}}} &=& \frac{G(V,0)}{A_{\textrm{j}}}\left[1+\frac{\pi^2k_{\textrm{B}}^2m_\textrm{e}d^2}{3\hbar^2(\phi_0-\frac{eV}{2})}T^2\right]\nonumber\\
&\text{ }& + J(V,0)\left[\frac{\pi^2k_{\textrm{B}}^2m_\textrm{e}ed^2}{6\hbar^2(\phi_0-\frac{eV}{2})^2}T^2\right],
\label{SimmonsJ(V,T)Derivative}
\end{eqnarray}
where $G(V,0)/A_{\textrm{j}}$ is defined by Eq. \eqref{SimmonsGV}, i.e.,
\begin{gather}
\frac{G}{A_\textrm{j}} = \frac{e^2}{8\pi hd^{2}}\Bigg\{\exp\left[-A\left(\phi_0-\frac{eV}{2}\right)^{\frac{1}{2}}\right]\left[A\left(\phi_0-\frac{eV}{2}\right)^{\frac{1}{2}}-2\right]\nonumber\\
+\exp\left[-A\left(\phi_0+\frac{eV}{2}\right)^{\frac{1}{2}}\right]\left[A\left(\phi_0+\frac{eV}{2}\right)^{\frac{1}{2}}-2\right]\Bigg\}
\label{SimmonsGVAppB}
\end{gather}
with $A=2d\sqrt{2m_{\text{e}}}/\hbar$ and $J(V,0)$ defined by Eq. \eqref{SimmonsJV}, i.e.,
\begin{eqnarray}
J(V,0) &=& \frac{e}{2\pi hd^{2}}\Bigg\{\left(\phi_0-\frac{eV}{2}\right)\exp\left[-A\left(\phi_0-\frac{eV}{2}\right)^{\frac{1}{2}}\right]\nonumber\\
&\text{ }& - \left(\phi_0+\frac{eV}{2}\right)\exp\left[-A\left(\phi_0+\frac{eV}{2}\right)^{\frac{1}{2}}\right]\Bigg\}. \nonumber
\label{SimmonsJVAppB}
\end{eqnarray}

By setting $V=0$ in Eq. \eqref{SimmonsJ(V,T)Derivative}, one gets
\begin{equation}
    \frac{G(0,T)}{A_{\textrm{j}}} = \frac{G(0,0)}{A_{\textrm{j}}}\left(1+\frac{\pi^2k_{\textrm{B}}^2m_\textrm{e}d^2}{3\hbar^2\phi_0}T^2\right),
\end{equation}
where 
\begin{equation*}
 \frac{G(0,0)}{A_{\textrm{j}}} = \frac{e^2\exp(-A\phi_0^{\frac{1}{2}})}{4\pi hd^2}( A\phi_0^{\frac{1}{2}}-2),  
\end{equation*}
which, by neglecting the term -2, reduces to Eq. \eqref{SimmonsG(T)a} in the main text
\begin{equation}
G(V=0,T) = G_{0,0}\left(1+\frac{T^2}{T_{0,0}^2}\right),
\label{SimmonsG(V,T)}
\end{equation}
with
\begin{equation}
T_{0,0}^2 = \frac{3\hbar^2\phi_0}{\pi^2k_{\textrm{B}}^2m_\textrm{e}d^2} \nonumber
\label{T_00AppB}
\end{equation}
and 
\begin{equation} 
G_{0,0} = \frac{e^2A_\textrm{j}\sqrt{2m_\textrm{e}\phi_0}}{h^2d}\textrm{exp}\biggr[\frac{-2d\sqrt{2m_\textrm{e}\phi_0}}{\hbar}\biggr] \nonumber
\label{G_00AppB}
\end{equation}
being the same as in the zero-temperature Simmons formula, Eq. \eqref{SimmonsSimpleGVatT=0}, i.e.,
\begin{equation}
G(V,T=0) = G_{0,0}\left(1+\frac{V^2}{V_{0,0}^2}\right).
\label{SimmonsSimpleGVatT=0AppB}
\end{equation}

Concerning the use of Simmons' temperature-dependent formulas, they have not been used in the literature as much as the zero-temperature ones. Many studies on tunnel barrier properties \cite{Wang2001, Rippard, Buchanan2002ApplPhysLett, Parkin2004, Koberidze} cite Simmons' model without specifying which formulas they use in the fitting. Ref. \onlinecite{Dorneles2003} cites Eq. \eqref{SimmonsGeneralizedJ(V,T)AppB} as a fit function in determining barrier heights and thicknesses from $I-V$ curves.  Eqs. \eqref{SimmonsJ(V,T)Derivative} and \eqref{SimmonsG(V,T)} are not shown in Simmons' papers \cite{SimmonsLowVoltage, SimmonsSimilarElectrodes, SimmonsDissimilarElectrodes, SimmonsThermal}, but Eq. \eqref{SimmonsG(V,T)} was presented in Ref. \onlinecite{Gloos2000} and reviewed in Ref. \onlinecite{Giazotto},  to be used as a thermometer. In addition to Eq. \eqref{SimmonsG(V,T)}, an attempt was made in Ref. \onlinecite{Gloos2000} to write a simple result for the conductance at both finite voltage and temperature, as a combination of Eqs. \eqref{SimmonsSimpleGVatT=0AppB} and \eqref{SimmonsG(V,T)}:
\begin{equation}
G(V,T) = G_{0,0}\left(1+\frac{V^2}{V_{0,0}^2}\right)\left(1+\frac{T^2}{T_{0,0}^2}\right).
\label{GloosG(V,T)}
\end{equation}
However, Eq. \eqref{SimmonsJ(V,T)Derivative} shows that Eq. \eqref{GloosG(V,T)} cannot be derived from Eq. \eqref{SimmonsGeneralizedJ(V,T)AppB} and is not correct, because the voltage dependence of the average barrier $\overline{\phi}(V)$ generates the second term in Eq. \eqref{SimmonsJ(V,T)Derivative}. 



\section{\label{J(V,T)} Derivation of the new $J(V,T)$ and $G(V,T)$ equations}

At nonzero temperatures, the elastic tunneling current density $J$ as a function of voltage $V$ and temperature $T$ is given\cite{Wolf, SimmonsThermal} by Eq. \eqref{WKBJ(V,T)Integral}, i.e., 
\begin{gather}
J(V,T) = Bk_{\textrm{B}}T\int_{0}^{E_m} D(E_x)\nonumber\\
\times \ln \Bigg\{\frac{1+\exp[(\mu-E_x)/(k_{\textrm{B}}T)]}{1+\exp[(\mu-E_x-eV)/(k_{\textrm{B}}T)]}\Bigg\}dE_x
\label{WKBJ(V,T)IntegralAppC}
\end{gather}
with
\begin{equation}
B=\frac{4\pi m_{\textrm{e}}e}{h^3}.\nonumber
\label{B'AppC}
\end{equation}
Using dimensionless variables $\alpha=A\overline{\phi}^{\frac{1}{2}}$, $b=\overline{\phi}/(k_\textrm{B}T)$, $\gamma=eV/\overline{\phi}$ and $\zeta=(E_x-\mu)/\overline{\phi}$, and working with Simmons' approximation for the tunneling probability, Eq. \eqref{D(Ex)},
\begin{eqnarray}
D(E_x) &\simeq& \exp\left[-A(\mu+\overline{\phi}-E_x)^{\frac{1}{2}}\right] \nonumber \\
&=& \exp\left[-\alpha(1-\zeta)^{\frac{1}{2}}\right], \nonumber
\label{D(Ex)AppC}
\end{eqnarray}
Eq. \eqref{WKBJ(V,T)IntegralAppC} becomes
\begin{equation}
J = B\overline{\phi}k_{\textrm{B}}T\int_{-\mu/\overline{\phi}}^{1}e^{-\alpha(1-\zeta)^{\frac{1}{2}}}\ln \left[\frac{1+e^{-b\zeta}}{1+e^{-b(\zeta+\gamma)}}\right]d\zeta,
\label{J(V,T)AfterChangeOfVariables}
\end{equation}
where the maximum energy of the tunneling electron $E_m = \mu+\overline{\phi}$ gives the upper limit of the integral $\zeta_m = 1$. 

Looking at the lower limit of the integral, it is the exponential function that suppresses the integrand for negative values of $\zeta$. Therefore, there are parameter ranges for $\alpha$ (barrier properties) where the integrand is negligible at $\zeta = -\mu/\overline{\phi}$ so that the lower limit can be extended to minus infinity (i.e. when the conduction band edge does not impact the current). We estimate that for typical barriers and metals this is indeed the case. For example, for a barrier with thickness $d = 1$ nm and average height $\overline{\phi}  = 1$ eV (which for a rectangular barrier with $ \overline{\phi} = \phi_0 - eV/2$ can for example correspond to the zero-bias height $\phi_0 = 1$ eV at $eV = 0$, or $\phi_0 = 2$ eV biased at $eV = \phi_0$) we get $\alpha = 10.2$, which translates to the integrand being negligible at $\zeta \approx -2.5$ and thus a condition for the Fermi energy $\mu >$ 2.5 eV, easily satisfied by most normal metals. Increasing $\overline{\phi}$ to 2 eV leads to $\alpha = 14.5$ and $\zeta_{min} \approx -1.5$, giving still an easily satisfiable condition $\mu >$ 3 eV.


Setting the lower integration limit to minus infinity and noting that as for typical barriers $k_BT \ll \overline{\phi}$ and thus $b \gg 1$ even at room temperature, the integrand is also negligible at the upper limit $\zeta = 1$, and we obtain by integrating by parts 
\begin{gather}
J = -B\overline{\phi}k_{\textrm{B}}T\int_{-\infty}^{1}\frac{2}{\alpha^2}e^{-\alpha(1-\zeta)^{\frac{1}{2}}}[\alpha(1-\zeta)^{\frac{1}{2}}+1] \nonumber \\
\times\frac{b(e^{-b(\zeta+\gamma)}-e^{-b\zeta})}{(1+e^{-b\zeta})(1+e^{-b(\zeta+\gamma)})}d\zeta.
\label{J(V,T)FirstPartialIntegration}
\end{gather}
Integrating by parts once more, we get
\begin{gather}
J = \frac{4b^2B\overline{\phi}k_{\textrm{B}}T}{\alpha^2}\int_{-\infty}^{1}\underbrace{e^{-\alpha(1-\zeta)^{\frac{1}{2}}}\left[\frac{3(1-\zeta)^{\frac{1}{2}}}{\alpha}+\frac{3}{\alpha^2}+1-\zeta\right]}_{=K(\zeta)} \nonumber \\
\times\left[\underbrace{\frac{e^{b\zeta}}{(e^{b\zeta}+1)^2}}_{=L(\zeta)}-\underbrace{\frac{e^{b(\zeta+\gamma)}}{(e^{b(\zeta+\gamma)}+1)^2}}_{=M(\zeta)}\right]d\zeta,
\label{J(V,T)SecondPartialIntegration}
\end{gather}
which can be written as a difference of two terms $J = J_1 -J_2$ in the form
\begin{gather}
J = \underbrace{\frac{4b^2B\overline{\phi}k_{\textrm{B}}T}{\alpha^2}\int_{-\infty}^{1}K(\zeta)L(\zeta)d\zeta}_{=J_1} \nonumber \\
-\underbrace{\frac{4b^2B\overline{\phi}k_{\textrm{B}}T}{\alpha^2}\int_{-\infty}^{1}K(\zeta)M(\zeta)d\zeta}_{=J_2},
\label{J(V,T)J1J2}
\end{gather}
with functions $K(\zeta)$, $L(\zeta)$ and $M(\zeta)$ defined in Eq. \eqref{J(V,T)SecondPartialIntegration}. The function $L$ has a narrow peak (width $\sim 1/b$) at $\zeta=0$, and the function $M$ similarly at $\zeta=-\gamma$, while $K$ is a more slowly varying function.  We therefore make a small error if we perform low-order Taylor expansions of $K$ around the peak positions. 

For $J_1$ the expansion around $\zeta = 0$ is then
\begin{equation}
J_1=\frac{4b^2B\overline{\phi}k_{\textrm{B}}T}{\alpha^2}\sum_{n=0}^{\infty}\frac{K^{(n)}(0)}{n!}\int_{-\infty}^{\infty}\zeta^n\frac{e^{b\zeta}}{(e^{b\zeta}+1)^2}d\zeta,
\label{J1SommerfeldExpansion}
\end{equation}
where $K^{(n)}(0)$ denotes the $n$:th derivative of $K$ at $\zeta = 0$, and the upper limit of integration was extended to plus infinity (no dependence $(1-\zeta)^{\frac{1}{2}}$ left), which gives a negligible error in the limit $b \gg 1$. 

Since $L$ is an even function that does not depend on the index $n$, the integrand of Eq. \eqref{J1SommerfeldExpansion} is odd when $n$ is odd, and the integral is then zero. If $b \gg 1$, we then make a small error by considering only the terms with $n=0$ and $n=2$ in Eq. \eqref{J1SommerfeldExpansion}, getting
\begin{equation}
J_1=\frac{4b^2B\overline{\phi}k_{\textrm{B}}T}{\alpha^2}\left[\frac{e^{-\alpha}}{b}\Big(\frac{3}{\alpha}+\frac{3}{\alpha^2}+1\Big)+\frac{\pi^2\alpha^2e^{-\alpha}}{24b^3}\right],
\label{J1Calculated}
\end{equation}
where we used the results 
\begin{eqnarray}
\int_{-\infty}^{\infty}\frac{e^{b\zeta}}{(e^{b\zeta}+1)^2}d\zeta &=& \frac{1}{b}, \nonumber\\
\int_{-\infty}^{\infty}\zeta^2\frac{e^{b\zeta}}{(e^{b\zeta}+1)^2}d\zeta &=& \frac{\pi^2}{3b^3}.
\label{integrals}
\end{eqnarray}

Similarly for the second term $J_2$, expanding $K$ as a Taylor series, but now around $\zeta=-\gamma$, allows $J_2$ to be written as
\begin{eqnarray}
J_2&=&\frac{4b^2B\overline{\phi}k_{\textrm{B}}T}{\alpha^2}\sum_{n=0}^{\infty}\frac{K^{(n)}(-\gamma)}{n!}\int_{-\infty}^{\infty}(\zeta+\gamma)^n\frac{e^{b(\zeta+\gamma)}}{(e^{b(\zeta+\gamma)}+1)^2}d\zeta \nonumber \\
&=&\frac{4b^2B\overline{\phi}k_{\textrm{B}}T}{\alpha^2}\sum_{n=0}^{\infty}\frac{K^{(n)}(-\gamma)}{n!}\int_{-\infty}^{\infty}(\zeta')^n\frac{e^{b\zeta'}}{(e^{b\zeta'}+1)^2}d\zeta'.
\label{J2SommerfeldExpansion}
\end{eqnarray}
 Again, the integrand of Eq. \eqref{J2SommerfeldExpansion} is odd when $n$ is odd, and the integral is then zero. If $b \gg 1$, we then again make a small error by considering only the terms with $n=0$ and $n=2$ in Eq. \eqref{J2SommerfeldExpansion}, getting with the help of the integrals of Eq. \eqref{integrals} 
\begin{gather}
J_2=\frac{4bB\overline{\phi}k_{\textrm{B}}T}{\alpha^2}e^{-\alpha(1+\gamma)^{\frac{1}{2}}}\left[\frac{3(1+\gamma)^{\frac{1}{2}}}{\alpha} \nonumber 
+\frac{3}{\alpha^2}+1+\gamma+\frac{\pi^2\alpha^2}{24b^2}\right]. 
\label{J2Calculated}
\end{gather}
The total current density is then
\begin{eqnarray}
J&=&J_1-J_2 \nonumber \\
&=&\frac{4bB\overline{\phi}k_{\textrm{B}}T}{\alpha^2}\Bigg\{e^{-\alpha}\Bigg[\frac{3}{\alpha}+\frac{3}{\alpha^2}+1+\frac{\pi^2\alpha^2}{24b^2}\Bigg] \nonumber \\
&-& e^{-\alpha(1+\gamma)^{\frac{1}{2}}}\Bigg[\frac{3(1+\gamma)^{\frac{1}{2}}}{\alpha}+\frac{3}{\alpha^2}  
 +1+\gamma+\frac{\pi^2\alpha^2}{24b^2}\Bigg]\Bigg\},
\label{JCalculated}
\end{eqnarray}
which is equivalent to the generalized result Eq. \eqref{ImprovedGeneralizedJ(V,T)} of the main text, i.e.,
\begin{gather}
J(V,T) = \frac{4B}{A^4}\Bigg\{\exp(-A{\overline{\phi}}^{\frac{1}{2}})\Big[A^2\overline{\phi}+3(A{\overline{\phi}}^{\frac{1}{2}}+1)+\frac{A^4\pi^2k_{\textrm{B}}^2T^2}{24}\Big]\nonumber\\
-\exp\Big[-A(\overline{\phi}+eV)^{\frac{1}{2}}\Big]\Bigg[A^2(\overline{\phi}+eV)\nonumber\\
+3\Big[A(\overline{\phi}+eV)^{\frac{1}{2}}+1\Big]+\frac{A^4\pi^2k_{\textrm{B}}^2T^2}{24}\Bigg]\Bigg\},
\label{ImprovedGeneralizedJ(V,T)AppC}
\end{gather}
with
\begin{equation}
A = \frac{4\pi\beta \Delta s\sqrt{2m_{\textrm{e}}}}{h}. \nonumber
\label{AAppC}
\end{equation}

To derive results for the trapezoidal barrier for bias voltages $|V| < \phi_1/e$, we make the approximation $\beta=1$ and substitute Eq. \eqref{overline_phi2}, i.e.,
\begin{equation}
\left\{ \begin{array}{ll}
\overline{\phi} = \phi_0 - \frac{eV}{2}\\
\Delta s = d_{\text{eff}} \equiv d \nonumber
\label{overline_phi2AppC}
\end{array}\right.
\end{equation}
into Eq. \eqref{ImprovedGeneralizedJ(V,T)AppC}, and get
\begin{gather}
J = \frac{4B}{A^4}\Bigg\{\exp\Big[-A\left(\phi_0-\frac{eV}{2}\right)^{\frac{1}{2}}\Big]\Bigg[A^2\left(\phi_0-\frac{eV}{2}\right)\nonumber\\
+3A\left(\phi_0-\frac{eV}{2}\right)^{\frac{1}{2}}+3+\frac{A^4\pi^2k_{\textrm{B}}^2T^2}{24}\Bigg]
-\exp\Big[-A\left(\phi_0+\frac{eV}{2}\right)^{\frac{1}{2}}\Big]\nonumber\\
\Bigg[A^2\left(\phi_0+\frac{eV}{2}\right)+3A\left(\phi_0+\frac{eV}{2}\right)^{\frac{1}{2}}+3+\frac{A^4\pi^2k_{\textrm{B}}^2T^2}{24}\Bigg]\Bigg\}
\label{ImprovedJ(V,T)AppC}
\end{gather} 
with $A=2d\sqrt{2m_{\text{e}}}/\hbar$. By differentiating Eq. \eqref{ImprovedJ(V,T)AppC} with respect to voltage we get Eq. \eqref{ImprovedG(V,T)}, i.e.,
\begin{gather}
\frac{G}{A_\textrm{j}} = \frac{e^2}{8\pi hd^{2}}\Bigg\{\exp\left[-A\left(\phi_0-\frac{eV}{2}\right)^{\frac{1}{2}}\right]\Bigg[A\left(\phi_0-\frac{eV}{2}\right)^{\frac{1}{2}}+1\nonumber\\
+\frac{A^3\pi^2k_{\textrm{B}}^2T^2}{24}\left(\phi_0-\frac{eV}{2}\right)^{-\frac{1}{2}}\Bigg]\nonumber\\
+\exp\left[-A\left(\phi_0+\frac{eV}{2}\right)^{\frac{1}{2}}\right]\Bigg[A\left(\phi_0+\frac{eV}{2}\right)^{\frac{1}{2}}+1\nonumber\\
+\frac{A^3\pi^2k_{\textrm{B}}^2T^2}{24}\left(\phi_0+\frac{eV}{2}\right)^{-\frac{1}{2}}\Bigg]\Bigg\}.
\label{ImprovedG(V,T)AppC}
\end{gather}
For low enough biases for which $eV < \phi_0$, Eq. \eqref{ImprovedG(V,T)AppC} can be described well by a second-order Taylor expansion with respect to $x=eV/\phi_0$, giving
\begin{gather*}
G = G_0\Bigg\{1+\frac{\pi^2}{24}\frac{(A\phi_0^{\frac{1}{2}})^3}{A\phi_0^{\frac{1}{2}}+1}\left(\frac{k_BT}{\phi_0}\right)^2\\
+ \frac{1}{32}\frac{(A\phi_0^{\frac{1}{2}})^3}{A\phi_0^{\frac{1}{2}}+1}\left[1+\frac{\pi^2}{24}(A^2\phi_0+3A\phi_0^{\frac{1}{2}}+3)\left(\frac{k_BT}{\phi_0}\right)^2\right]\left(\frac{eV}{\phi_0}\right)^2\Bigg\},
\end{gather*}
which is equivalent to Eq. \eqref{ParabolicApproximationOfImprovedG(V,T)}, i.e.,
\begin{gather}
G(V,T) = \underbrace{\frac{G_0}{V_0^2}\Bigg[1+\underbrace{\frac{4\pi^2k_{\textrm{B}}^2T^2}{3}(1+C)\left(\frac{1}{(eV_0)^2}+\frac{3}{32\phi_0^2C}\right)}_{=a_T}\Bigg]}_{=a}V^2\nonumber\\
+\underbrace{G_0\left(1+\underbrace{\frac{T^2}{T_0^2}}_{=c_T}\right)}_{=c}
\label{ParabolicApproximationOfImprovedG(V,T)AppC}
\end{gather}
where the corrected parameters for zero-temperature zero-bias conductance $G_0$, the zero-temperature voltage curvature $V_0$, and the temperature curvature $T_0$ are defined as
\begin{equation} 
G_0 = G_{0,0}(1+C), \nonumber
\label{G_0AppC}
\end{equation}
\begin{equation}
V_0^2 = V_{0,0}^2(1+C), \nonumber
\label{V_0AppC}
\end{equation}
and
\begin{equation}
T_0^2 = T_{0,0}^2(1+C), \nonumber
\label{T_0AppC}
\end{equation}
where $C$ is a dimensionless correction factor 
\begin{equation}
C = \frac{\hbar}{2d\sqrt{2m_{\textrm{e}}\phi_0}}, \nonumber
\label{CAppC}
\end{equation}
and the original Simmons parameters are 
\begin{equation} 
G_{0,0} = \frac{e^2A_\textrm{j}\sqrt{2m_\textrm{e}\phi_0}}{h^2d}\textrm{exp}\biggr[\frac{-2d\sqrt{2m_\textrm{e}\phi_0}}{\hbar}\biggr], \nonumber
\label{G_00AppC}
\end{equation}
\begin{equation}
V_{0,0}^2 = \frac{4\hbar^2\phi_0}{e^2m_\textrm{e}d^2}, \nonumber
\label{V_00AppC}
\end{equation}
and
\begin{equation}
T_{0,0}^2 = \frac{3\hbar^2\phi_0}{\pi^2k_{\textrm{B}}^2m_\textrm{e}d^2}. \nonumber
\label{T_00AppC}
\end{equation}

\begin{figure}[]
\includegraphics[width=1.0\linewidth]{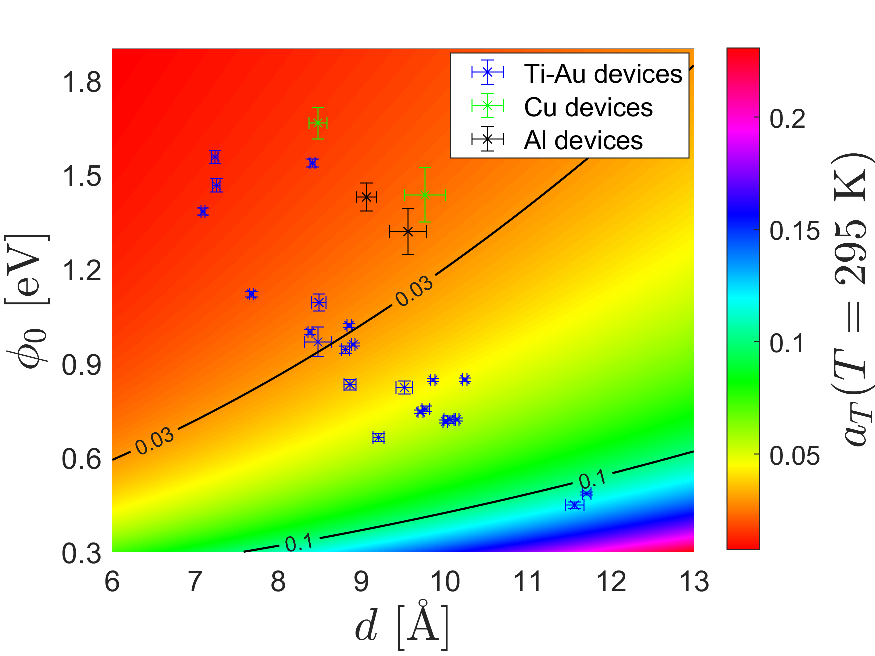}
\caption{\label{GVFitParametersRTZBWithaTAndNewTheory} The quantity $a_T$ at $T=295\text{ K}$ (color scale) with respect to $d$ and $\phi_0$ together with fit parameters $d$ and $\phi_0$ obtained for Ti-Au (blue points), Cu (green points) and Al devices (black points) by fitting Eq. \eqref{ParabolicApproximationOfImprovedG(V,T)AppC} to experimental $G-V$ data taken at room temperature.}
\end{figure}
\begin{figure}[]
\includegraphics[width=1.0\linewidth]{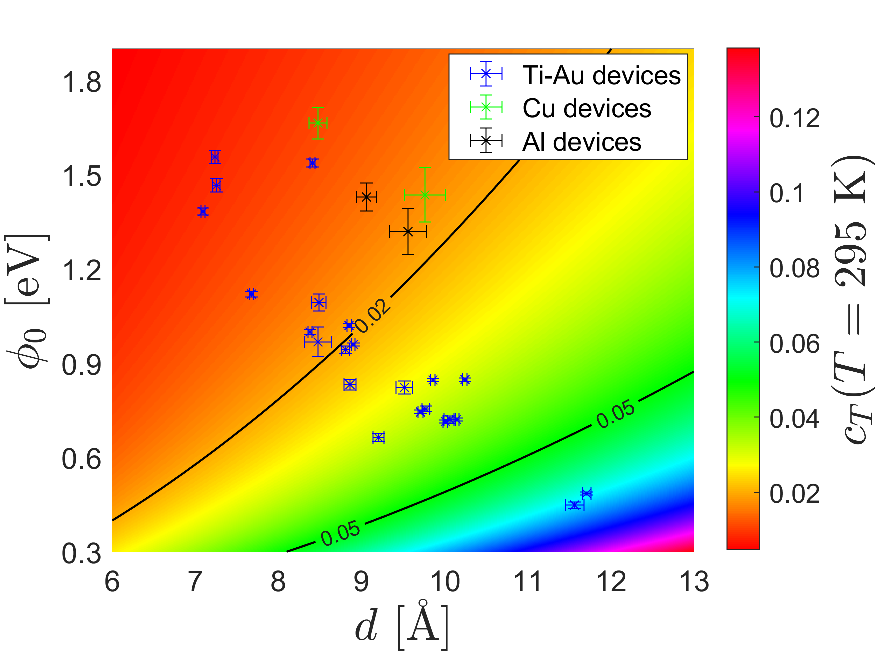}
\caption{\label{GVFitParametersRTZBWithcTAndNewTheory} The quantity $c_T$ at $T=295\text{ K}$ (color scale) with respect to $d$ and $\phi_0$ together with fit parameters $d$ and $\phi_0$ obtained for Ti-Au (blue points), Cu (green points) and Al devices (black points) by fitting Eq. \eqref{ParabolicApproximationOfImprovedG(V,T)AppC} to experimental $G-V$ data taken at room temperature.}
\end{figure}

Eq. \eqref{ParabolicApproximationOfImprovedG(V,T)AppC} describes a parabola with respect to $V$ of the form $G(V)=aV^2+c$, where the coefficients $a$ and $c$ are both temperature dependent, as can be seen from Eq. \eqref{ParabolicApproximationOfImprovedG(V,T)AppC}. The temperature-dependent parts of $a$ and $c$ are denoted in Eq. \eqref{ParabolicApproximationOfImprovedG(V,T)AppC} by the dimensionless quantities $a_T$ and $c_T$, respectively, which describe the relative strength of the finite-temperature corrections. Figs. \ref{GVFitParametersRTZBWithaTAndNewTheory} and \ref{GVFitParametersRTZBWithcTAndNewTheory} show $a_T$ and $c_T$ at $T=295\text{ K}$ as a function of $d$ and $\phi_0$, together with parameters $d$ and $\phi_0$ obtained for representative Ti-Au, Cu and Al devices by fitting Eq. \eqref{ParabolicApproximationOfImprovedG(V,T)AppC} to experimental $G-V$ data taken at room temperature. The fit parameters $d$ and $\phi_0$ shown in Figs. \ref{GVFitParametersRTZBWithaTAndNewTheory} and \ref{GVFitParametersRTZBWithcTAndNewTheory} are the same as those shown in Fig. \ref{GVFitParametersRTZBWithCAndNewTheory}. For some devices, $a_T>0.1$ and $c_T>0.05$, indicating that the temperature correction is significant at room temperature.

\section{\label{J(V,0)} Additional derivation of the new $J-V$ and $G-V$ equations at $T=0$}

We begin the derivation of the new $J-V$ and $G-V$ equations at $T=0$ from the same starting point as Simmons, Eqs. \eqref{WKBJ(V)Integral} and \eqref{D(Ex)}, rewritten below for clarity as Eqs. \eqref{WKBJ(V)IntegralAppD} and  \eqref{D(Ex)AppD}, where Eq. \eqref{D(Ex)AppD} is Simmons' approximation for the tunneling probability derived in Appendix \ref{SimmonsJVGV}. 

At zero temperature, the elastic tunneling current density $J$ as a function of voltage $V$ is described by Eq. \eqref{WKBJ(V)Integral}, i.e., 
\begin{gather}
J = B\Big[eV\int_{0}^{\mu-eV} D(E_x)dE_x
+\int_{\mu-eV}^{\mu}(\mu-E_x)D(E_x)dE_x\Big]
\label{WKBJ(V)IntegralAppD}
\end{gather}
with
\begin{equation}
B=\frac{4\pi m_{\textrm{e}}e}{h^3}, \nonumber
\label{B'AppD}
\end{equation}
and Simmons' approximation for the tunneling probability is Eq. \eqref{D(Ex)}, i.e.,
\begin{equation}
D(E_x) \simeq \exp\left[-A(\mu+\overline{\phi}-E_x)^{\frac{1}{2}}\right]
\label{D(Ex)AppD}
\end{equation}
with
\begin{equation}
A = \frac{4\pi\beta \Delta s\sqrt{2m_{\textrm{e}}}}{h}. \nonumber
\label{AAppD}
\end{equation}
By substituting Eq. \eqref{D(Ex)AppD} into Eq. \eqref{WKBJ(V)IntegralAppD}, 
we get
\begin{gather}
J = B\Bigg\{eV\int_{0}^{\mu-eV} \exp[-A(\mu+\overline{\phi}-E_x)^{\frac{1}{2}}]dE_x\nonumber\\
+\int_{\mu-eV}^{\mu}(\mu-E_x)\exp[-A(\mu+\overline{\phi}-E_x)^{\frac{1}{2}}]dE_x\Bigg\}.
\label{SimilarSimmonsEq12AppD}
\end{gather}

To allow easier comparison to Simmons' results, we write Eq. \eqref{SimilarSimmonsEq12AppD} in the form
\begin{gather}
J = B\Bigg\{eV\int_{0}^{\mu-eV} \exp[-A(\mu+\overline{\phi}-E_x)^{\frac{1}{2}}]dE_x\nonumber\\
-\overline{\phi}\int_{\mu-eV}^{\mu}\exp[-A(\mu+\overline{\phi}-E_x)^{\frac{1}{2}}]dE_x\nonumber\\
+\int_{\mu-eV}^{\mu}(\mu+\overline{\phi}-E_x)\exp[-A(\mu+\overline{\phi}-E_x)^{\frac{1}{2}}]dE_x\Bigg\}
\label{SimilarSimmonsEq13AppD}
\end{gather}
which is Eq. (13) of Ref. \onlinecite{SimmonsSimilarElectrodes}. The first integral  yields for the first term in Eq. \eqref{SimilarSimmonsEq13AppD}
\begin{gather}
(2BeV/A^2)\{[A({\overline{\phi}}+eV)^{\frac{1}{2}}+1]\exp[-A({\overline{\phi}}+eV)^{\frac{1}{2}}]\nonumber\\
-[A(\overline{\phi}+\mu)^{\frac{1}{2}}+1]\exp[-A(\overline{\phi}+\mu)^{\frac{1}{2}}]\}.
\label{SimilarSimmonsEq14AppD}
\end{gather}
If we consider the limit $\mu > eV$, which is satisfied for regular metallic electrodes up to high voltages $\sim 5$ V,  the second term in the braces in Eq. \eqref{SimilarSimmonsEq14AppD} is exponentially smaller than the first term and can be neglected. 

The second integral gives for the second term in Eq. \eqref{SimilarSimmonsEq13AppD}
\begin{gather}
(2B\overline{\phi}/A^2)\{[A({\overline{\phi}}+eV)^{\frac{1}{2}}+1]\exp[-A({\overline{\phi}}+eV)^{\frac{1}{2}}]\nonumber\\
-[A\overline{\phi}^{\frac{1}{2}}+1]\exp[-A\overline{\phi}^{\frac{1}{2}}]\},
\label{SimilarSimmonsEq16AppD}
\end{gather}
and by defining an integration variable $z=(\mu+\overline{\phi}-E_x)^{\frac{1}{2}}$, the third integral  gives for the third term in Eq. \eqref{SimilarSimmonsEq13AppD}
\begin{eqnarray}
&-& 2B\int_{(\overline{\phi}+eV)^{\frac{1}{2}}}^{\overline{\phi}^{\frac{1}{2}}}z^3\exp(-Az)dz \nonumber\\
&=& 2B\exp(-Az)\Big(\frac{z^3}{A}+\frac{3z^2}{A^2}+\frac{6z}{A^3}+\frac{6}{A^4}\Big) \bigg|_{(\overline{\phi}+eV)^{\frac{1}{2}}}^{\overline{\phi}^{\frac{1}{2}}} \nonumber\\
&=& (2B/A^4)\exp(-A\overline{\phi}^{\frac{1}{2}})[A^3\overline{\phi}^{\frac{3}{2}}+3A^2\overline{\phi}+6A\overline{\phi}^{\frac{1}{2}}+6]\nonumber\\
&\text{ }& - (2B/A^4)\exp[-A(\overline{\phi}+eV)^{\frac{1}{2}}][A^3(\overline{\phi}+eV)^{\frac{3}{2}}\nonumber\\
&\text{ }& + 3A^2(\overline{\phi}+eV)+6A(\overline{\phi}+eV)^{\frac{1}{2}}+6].
\label{WKBJ(V)IntegralThirdIntegralTrickD}
\end{eqnarray}

Summing Eq. \eqref{SimilarSimmonsEq14AppD} without its insignificant latter term $-[A(\overline{\phi}+\mu)^{\frac{1}{2}}+1]\exp[-A(\overline{\phi}+\mu)^{\frac{1}{2}}]$, Eq. \eqref{SimilarSimmonsEq16AppD} and Eq. \eqref{WKBJ(V)IntegralThirdIntegralTrickD}, we finally get
\begin{gather}
J = \frac{4B}{A^4}\Bigg\{\exp(-A{\overline{\phi}}^{\frac{1}{2}})\Big[A^2\overline{\phi}+3(A{\overline{\phi}}^{\frac{1}{2}}+1)\Big]\nonumber\\
-\exp\Big[-A(\overline{\phi}+eV)^{\frac{1}{2}}\Big]\Bigg[A^2(\overline{\phi}+eV)\nonumber\\
+3\Big[A(\overline{\phi}+eV)^{\frac{1}{2}}+1\Big]\Bigg]\Bigg\}.
\label{ImprovedGeneralizedJV}
\end{gather}

Compared to Simmons' original result, Eq. \eqref{SimmonsGeneralizedJV} (Eq. 20 of Ref. \onlinecite{SimmonsSimilarElectrodes}), Eq. \eqref{ImprovedGeneralizedJV} has two additional terms multiplying the exponentials, $3(A{\overline{\phi}}^{\frac{1}{2}}+1)$ and $3\Big[A(\overline{\phi}+eV)^{\frac{1}{2}}+1\Big]$, which result from the fact that instead of neglecting the terms $6z/A^3$ and $6/A^4$ as Simmons does, they are taken into account in Eq. \eqref{WKBJ(V)IntegralThirdIntegralTrickD}. We remark that Eq. \eqref{ImprovedGeneralizedJV} agrees with the more general finite-temperature result Eq. \eqref{ImprovedGeneralizedJ(V,T)} at $T=0$. 

To derive results for the trapezoidal barrier for bias voltages $|V| < \phi_1/e$, we make the approximation $\beta=1$ and substitute Eq. \eqref{overline_phi2}, i.e.,
\begin{equation}
\left\{ \begin{array}{ll}
\overline{\phi} = \phi_0 - \frac{eV}{2}\\
\Delta s = d \nonumber
\label{overline_phi2AppD}
\end{array}\right.
\end{equation}
into Eq. \eqref{ImprovedGeneralizedJV}. We get
\begin{gather}
J = \frac{4B}{A^4}\Bigg\{\exp\Bigg[-A\left(\phi_0-\frac{eV}{2}\right)^{\frac{1}{2}}\Bigg]\Bigg[A^2\left(\phi_0-\frac{eV}{2}\right)\nonumber\\
+3A\left(\phi_0-\frac{eV}{2}\right)^{\frac{1}{2}}+3\Bigg]
-\exp\Bigg[-A\left(\phi_0+\frac{eV}{2}\right)^{\frac{1}{2}}\Bigg]\nonumber\\
\Bigg[A^2\left(\phi_0+\frac{eV}{2}\right)+3A\left(\phi_0+\frac{eV}{2}\right)^{\frac{1}{2}}+3\Bigg]\Bigg\}
\label{ImprovedJV}
\end{gather}
with $A=2d\sqrt{2m_{\text{e}}}/\hbar$. Eq. \eqref{ImprovedJV} agrees with Eq. \eqref{ImprovedJ(V,T)AppC} in the limit $T=0$. By differentiating Eq. \eqref{ImprovedJV} with respect to voltage, we get
\begin{gather}
\frac{G}{A_\textrm{j}} = \frac{e^2}{8\pi hd^{2}}\Bigg\{\exp\left[-A\left(\phi_0-\frac{eV}{2}\right)^{\frac{1}{2}}\right]\left[A\left(\phi_0-\frac{eV}{2}\right)^{\frac{1}{2}}+1\right]\nonumber\\
+\exp\left[-A\left(\phi_0+\frac{eV}{2}\right)^{\frac{1}{2}}\right]\left[A\left(\phi_0+\frac{eV}{2}\right)^{\frac{1}{2}}+1\right]\Bigg\},
\label{ImprovedGV}
\end{gather}
which agrees with Eq. \eqref{ImprovedG(V,T)} in the limit $T=0$. 

For low enough biases for which $eV < \phi_0$, Eq. \eqref{ImprovedGV} can be described well by a second-order Taylor expansion with respect to $x=eV/\phi_0$, giving 
\begin{equation*}
G = G_0\left[1+\frac{1}{32}\frac{(A\phi_0^{\frac{1}{2}})^3}{A\phi_0^{\frac{1}{2}}+1}\left(\frac{eV}{\phi_0}\right)^2\right],
\end{equation*}
which is equivalent to Eq. \eqref{SimmonsG(V)}, i.e.
\begin{equation}
G(V,T=0) = G_0\left(1+\frac{V^2}{V_0^2}\right),
\label{SimmonsImprovedGVatT=0}
\end{equation}
where the corrected parameters for zero-bias conductance $G_0$ and voltage curvature $V_0$ are defined as
\begin{equation} 
G_0 = G_{0,0}(1+C), \nonumber
\label{G_0AppC}
\end{equation}
\begin{equation}
V_0^2 = V_{0,0}^2(1+C), \nonumber
\label{V_0AppC}
\end{equation}
with $C$ a dimensionless correction factor 
\begin{equation}
C = \frac{\hbar}{2d\sqrt{2m_{\textrm{e}}\phi_0}}, \nonumber
\label{CAppC}
\end{equation}
and the original Simmons parameters 
\begin{equation} 
G_{0,0} = \frac{e^2A_\textrm{j}\sqrt{2m_\textrm{e}\phi_0}}{h^2d}\textrm{exp}\biggr[\frac{-2d\sqrt{2m_\textrm{e}\phi_0}}{\hbar}\biggr], \nonumber
\label{G_00AppC}
\end{equation}
\begin{equation}
V_{0,0}^2 = \frac{4\hbar^2\phi_0}{e^2m_\textrm{e}d^2}. \nonumber
\label{V_00AppC}
\end{equation}


\section{\label{WKBGV} Derivation of the $G-V$ equation [Eq. \eqref{WKB G(V,0)}] from the WKB approximation}

At zero temperature, the elastic tunneling current density $J$ as a function of voltage $V$ is described by Eq. \eqref{WKBJ(V)Integral}, i.e., 
\begin{gather}
J = B\Big[eV\int_{0}^{\mu-eV} D(E_x)dE_x
+\int_{\mu-eV}^{\mu}(\mu-E_x)D(E_x)dE_x\Big]
\label{WKBJ(V)IntegralAppE}
\end{gather}
with
\begin{equation}
B=\frac{4\pi m_{\textrm{e}}e}{h^3}. \nonumber
\label{B'AppE}
\end{equation}

Here, we consider only a rectangular tunnel barrier with $\phi_0=\phi_1=\phi_2$, which has physical thickness $d_\textrm{phys}$. A bias voltage $V$ tilts the barrier, making its shape trapezoidal, $\phi(x,V)=\phi_0-eVx/d_\textrm{phys}$, assuming $V<\phi_0/e$. The WKB tunneling probability, Eq. \eqref{D(Ex)SecondForm}, can then be integrated analytically.
By writing $s_1=0$, $s_2=d_\textrm{phys}$, $g=\mu+\phi_0-E_x$, $k=eV/d_\textrm{phys}$, Eq. \eqref{D(Ex)SecondForm} becomes
\begin{eqnarray}
D(E_x) &=& \exp\left[-\frac{4\pi\sqrt{2m^{*}}}{h}\int_{s_1}^{s_2}(\mu+\phi(x)-E_x)^{\frac{1}{2}}dx \right] \nonumber \\
&=& \exp\left[-\frac{4\pi\sqrt{2m^{*}}}{h}\int_{0}^{d_\textrm{phys}}(g-kx)^{\frac{1}{2}}dx \right] \nonumber \\
&=& \exp\left[-\frac{4\pi\sqrt{2m^{*}}}{h}\frac{2}{3k}\Big(g^{\frac{3}{2}}-(g-kd_\textrm{phys})^{\frac{3}{2}}\Big)\right], \nonumber
\label{D(V,Ex)Derivation}
\end{eqnarray}
which is equivalent to Eq. \eqref{D(eV,Ex)}, i.e., 
\begin{gather}
D_{\textrm{WKB}}(V, E_x) = \exp\Bigg\{-\frac{A'}{eV}\Big[(\mu+\phi_0-E_x)^{\frac{3}{2}} \nonumber\\ 
-(\mu+\phi_0-E_x-eV)^{\frac{3}{2}}\Big]\Bigg\}
\label{D(eV,Ex)AppE}
\end{gather}
with
\begin{equation}
A' = \frac{8\pi d\sqrt{2m_{\textrm{e}}}}{3h},\nonumber
\label{A'}
\end{equation}
where $d \equiv d_\textrm{eff}$ now denotes the effective thickness defined in Eq. \eqref{dphys} as $d_\textrm{eff} = d_\textrm{phys}\sqrt{m^{*}/m_\textrm{e}}$.

Substitution of Eq. \eqref{D(eV,Ex)AppE} into Eq. \eqref{WKBJ(V)IntegralAppE} yields
\begin{gather}
J = \underbrace{B\int_{0}^{\mu-eV} \underbrace{D_{\textrm{WKB}}(V,E_x)eV}_{=p(V,E_x)}dE_x}_{=J_A} \nonumber \\
+B\int_{\mu-eV}^{\mu}D_{\textrm{WKB}}(V,E_x)(\mu-E_x)dE_x.
\label{WKBJ(V)IntegralMoreDetails}
\end{gather}
To obtain the differential conductance $G$, we need to differentiate Eq. \eqref{WKBJ(V)IntegralMoreDetails} with respect to voltage, which we accomplish by applying the Leibniz integral rule. The voltage derivative of the term $J_A$ is then given by
\begin{eqnarray}
\frac{dJ_A}{dV}&=&B\frac{d}{dV}\left[\int_{a(V)}^{b(V)}p(V,E_x)dE_x\right]\nonumber\\
&=&B\Big[p(V,b(V))\cdot\frac{db(V)}{dV} 
-p(V,a(V))\cdot\frac{da(V)}{dV}\nonumber \\
&\text{ }& + \int_{a(V)}^{b(V)}\frac{\partial{p(V,E_x)}}{\partial{V}}dE_x\Big]
\label{Leibniz}
\end{eqnarray}
with $a(V)=0$ and $b(V)=\mu-eV$, with the second term in Eq. \eqref{WKBJ(V)IntegralMoreDetails} integrated in the same way by substituting $p(V,E_x) = D_{\textrm{WKB}}(V,E_x)(\mu-E_x)$, $a(V)=\mu-eV$ and $b(V)=\mu$ in Eq. \eqref{Leibniz}. Summing the two contributions, we obtain 
\begin{gather}
\frac{G}{A_{\textrm{j}}}=-Be^2VD_{\textrm{WKB}}(V,\mu-eV) \nonumber \\
+B\int_0^{\mu-eV}\frac{\partial}{\partial{V}}\Big[D_{\textrm{WKB}}(V,E_x)eV\Big]dE_x \nonumber \\
+Be^2VD_{\textrm{WKB}}(V,\mu-eV) \nonumber \\
+B\int_{\mu-eV}^{\mu}\frac{\partial}{\partial{V}}\Big[D_{\textrm{WKB}}(V,E_x)\Big](\mu-E_x)dE_x,
\label{WKB G(V,0) more details}
\end{gather}
which, after cancellation of the first and third terms and differentiation, is equivalent to Eq. \eqref{WKB G(V,0)}, i.e.,
\begin{gather}
\frac{G}{A_\textrm{j}} = Be\Bigg(\int_{0}^{\mu-eV}D_{\textrm{WKB}}(V,E_x)\Bigg\{1+\frac{A'}{eV}\Bigg[(\mu+\phi_0-E_x)^{\frac{3}{2}}\nonumber\\
-(\mu+\phi_0-E_x-eV)^{\frac{3}{2}}-\frac{3eV}{2}(\mu+\phi_0-E_x-eV)^{\frac{1}{2}}\Bigg]\Bigg\}dE_x\nonumber\\
+\int_{\mu-eV}^{\mu}(\mu-E_x)D_{\textrm{WKB}}(V,E_x)\frac{A'}{(eV)^2}\Bigg[(\mu+\phi_0-E_x)^{\frac{3}{2}}\nonumber\\
-(\mu+\phi_0-E_x-eV)^{\frac{3}{2}}-\frac{3eV}{2}(\mu+\phi_0-E_x-eV)^{\frac{1}{2}}\Bigg]dE_x\Bigg).
\label{WKB G(V,0)AppE}
\end{gather}

\section{\label{CalculationOfBarrierParameters} Calculation of barrier parameters $d$ and $\phi_0$ and their errors after fitting Eq. \eqref{ParabolicApproximationOfImprovedG(V,T)} to experimental $G-V$ data}

The goal of this Appendix is to show that barrier parameters $d$ and $\phi_0$ can be determined by  fitting Eq. \eqref{ParabolicApproximationOfImprovedG(V,T)} to experimental $G-V$ data of a tunnel junction. We rewrite Eq. \eqref{ParabolicApproximationOfImprovedG(V,T)} here for convenience:
\begin{gather}
G(V,T) = \underbrace{\frac{G_0}{V_0^2}\Bigg[1+\frac{4\pi^2k_{\textrm{B}}^2T^2}{3}(1+C)\left(\frac{1}{(eV_0)^2}+\frac{3}{32\phi_0^2C}\right)\Bigg]}_{=a}V^2\nonumber\\
+\underbrace{G_0\left(1+\frac{T^2}{T_0^2}\right)}_{=c}
\label{ParabolicApproximationOfImprovedG(V,T)AppF}
\end{gather}
with
\begin{equation} 
G_0 = G_{0,0}(1+C), \nonumber
\label{G_0AppF}
\end{equation}
\begin{equation}
V_0^2 = V_{0,0}^2(1+C), \nonumber
\label{V_0AppF}
\end{equation}
\begin{equation}
T_0^2 = T_{0,0}^2(1+C), \nonumber
\label{T_0AppF}
\end{equation}
\begin{equation} 
G_{0,0} = \frac{e^2A_\textrm{j}\sqrt{2m_\textrm{e}\phi_0}}{h^2d}\textrm{exp}\biggr[\frac{-2d\sqrt{2m_\textrm{e}\phi_0}}{\hbar}\biggr], \nonumber
\label{G_00AppF}
\end{equation}
\begin{equation}
V_{0,0}^2 = \frac{4\hbar^2\phi_0}{e^2m_\textrm{e}d^2}, \nonumber
\label{V_00AppF}
\end{equation}
\begin{equation}
T_{0,0}^2 = \frac{3\hbar^2\phi_0}{\pi^2k_{\textrm{B}}^2m_\textrm{e}d^2}, \nonumber
\label{T_00AppF}
\end{equation}
and
\begin{equation}
C = \frac{\hbar}{2d\sqrt{2m_{\textrm{e}}\phi_0}}, \nonumber
\label{CAppF}
\end{equation}
showing explicitly that Eq. \eqref{ParabolicApproximationOfImprovedG(V,T)AppF} is a parabola with respect to $V$ with the form $G(V)=aV^2+c$. To fit Eq. \eqref{ParabolicApproximationOfImprovedG(V,T)AppF} to data, we use a linear least-squares algorithm, which gives the coefficients $a$ and $c$ and their 95 \% confidence intervals. We define the errors of $a$ and $c$, $\Delta a$ and $\Delta c$, respectively, as half of the 95 \% confidence interval lengths. 

\begin{figure}[]
\includegraphics[width=1.0\linewidth]{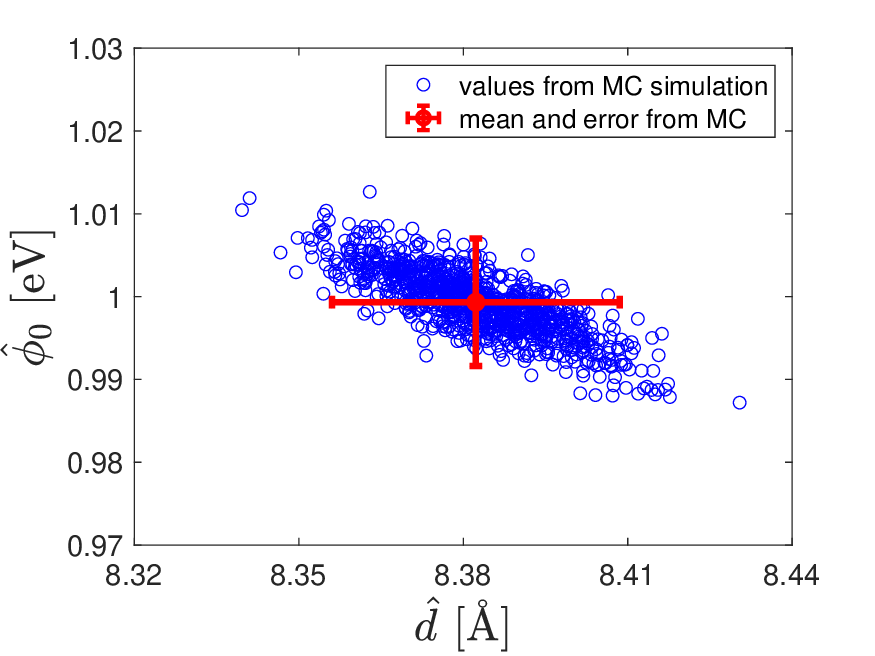}
\caption{\label{ErrorsOfdAndfii0LargeTiAuDeviceRT} Blue circles: 1000 values of $\hat{d}$ and $\hat{\phi}_0$ from a Monte Carlo simulation for the fit in Fig. \ref{GVLargeTiAuRTOneJunction}. Red: Mean and error based on standard deviations of $\hat{d}$ and $\hat{\phi}_0$.}
\end{figure}

After obtaining the values of $a$ and $c$ we can calculate $d$ and $\phi_0$ by numerically solving the nonlinear system 
\begin{equation}
\left\{ \begin{array}{ll}
F=a-\frac{e^4m_{\text{e}}^{3/2}A_{\text{j}}d}{4\hbar^2h^2}\sqrt{\frac{2}{\phi_0}}\exp{\Big(-\frac{2d\sqrt{2m_{\text{e}}\phi_0}}{\hbar}\Big)} \\
\times\Bigg[1+\frac{(\pi k_{\textrm{B}}T)^2}{\phi_0}\Big(\frac{m_{\text{e}}d^2}{3\hbar^2}+\frac{d}{2\hbar}\sqrt{\frac{m_{\text{e}}}{2\phi_0}}+\frac{1}{8\phi_0}\Big)\Bigg]=0\\
H=c-\frac{e^2A_{\text{j}}\sqrt{2m_{\text{e}}\phi_0}}{h^2d}\exp{\Big(-\frac{2d\sqrt{2m_{\text{e}}\phi_0}}{\hbar}\Big)} \\
\times\Bigg[\frac{\pi^2m_{\text{e}}d^2k_{\textrm{B}}^2T^2}{3\hbar^2\phi_0}+1+\frac{\hbar}{2d\sqrt{2m_{\text{e}}\phi_0}}\Bigg]=0
\end{array}\right.
\label{SystemForSolvingdAndfii0}
\end{equation}
that follows from Eq. \eqref{ParabolicApproximationOfImprovedG(V,T)AppF}, using fixed-point iteration. We get the errors of $d$ and $\phi_0$ from a Monte Carlo (MC) simulation, an example of which is shown in Fig. \ref{ErrorsOfdAndfii0LargeTiAuDeviceRT}. We define normally distributed random variables $\hat{a}\sim N(a,\sigma_a)$, $\hat{c}\sim N(c,\sigma_c)$, and $\hat{A}_{\text{j}}\sim N(A_{\text{j}},\sigma_{A_{\text{j}}})$. The quantities $\sigma_a$, $\sigma_c$ and $\sigma_{A_{\text{j}}}$ are the standard deviations of $\hat{a}$, $\hat{c}$ 
and $\hat{A}_{\text{j}}$, respectively, given by the least-squares fitting algorithm for $a$ and $c$ and the known experimental error for $A_{\text{j}}$. For the normal distribution, the standard deviation is the length of the 95 \% confidence interval divided by 3.92, or the error divided by 1.96. The simulation follows by drawing one value of $\hat{a}$, $\hat{c}$ and $\hat{A}_{\text{j}}$ and calculating the values of random variables $\hat{d}$ and $\hat{\phi}_0$, by solving Eq. \eqref{SystemForSolvingdAndfii0}. Now $\hat{d}$ and $\hat{\phi}_0$ are also normally distributed and their errors are equal to 1.96 times their standard deviations, i.e., $\Delta d=1.96\sigma_d$ and $\Delta \phi_0=1.96\sigma_{\phi_0}$. We repeat until we have drawn 1000 values of ($\hat{a}$, $\hat{c}$, $\hat{A}_{\text{j}}$) and calculated the values of $\hat{d}$ and $\hat{\phi}_0$ 1000 times.
\begin{figure}[]
\includegraphics[width=1.0\linewidth]{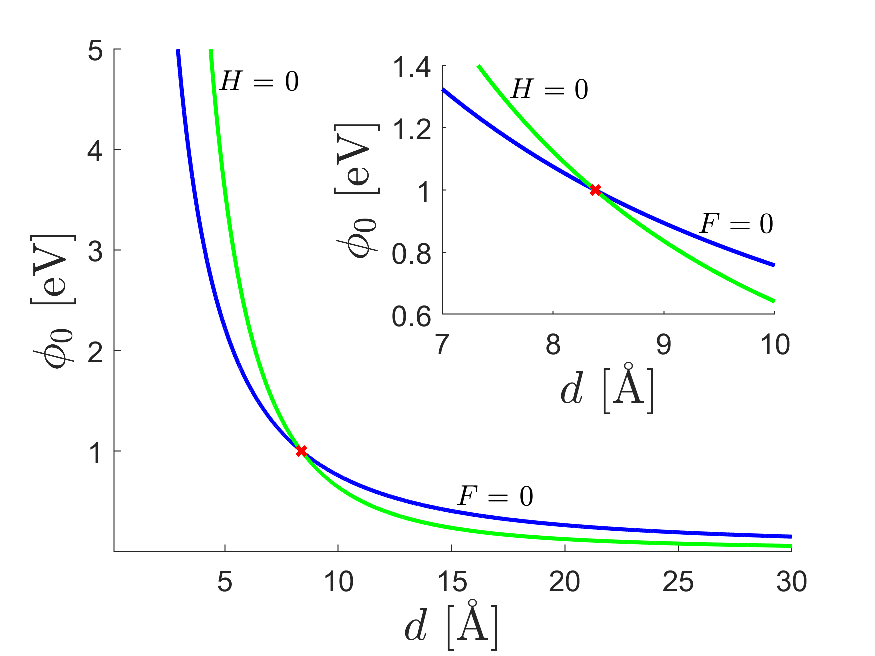}
\caption{\label{UniquenessOfdAndfii0LargeTiAuDeviceRT} Contours $F=0$ (blue curve) and $H=0$ (green curve) for the fit in Fig. \ref{GVLargeTiAuRTOneJunction} in the parameter space $\{0.1\textrm{ Å}\leq d\leq 30\textrm{ Å};\text{ }10^{-4}\textrm{ eV}\leq \phi_0\leq 5\textrm{ eV}\}$, intersecting at one single point (red cross) and showing there is only one reasonable solution for $d$ and $\phi_0$. Inset: detail showing the intersection point of the contours.}
\end{figure}

For example, for the data and fit shown in Fig. \ref{GVLargeTiAuRTOneJunction}, we obtain 
\begin{equation}
\left\{ \begin{array}{ll}
d \approx (8.38\pm0.03)\text{ Å}\\
\phi_0 \approx (1.00\pm0.01)\text{ eV}
\end{array}\right.
\label{ValuesOfdAndfii0}
\end{equation}
with $A_{\text{j}}=(0.52\pm0.01)\text{ }\mu\text{m}^2$ and $T=295\text{ K}$. 

Finally, in Fig. \ref{UniquenessOfdAndfii0LargeTiAuDeviceRT} we show that the solution of Eq. \eqref{SystemForSolvingdAndfii0} given by Eq. \eqref{ValuesOfdAndfii0} is unique. The contours $F=0$ and $H=0$ intersect at one single point in the parameter space $\{0.1\textrm{ Å}\leq d\leq 30\textrm{ Å};\text{ }10^{-4}\textrm{ eV}\leq \phi_0\leq 5\textrm{ eV}\}$, and the intersection point is given by Eq. \eqref{ValuesOfdAndfii0}. There might be other solutions for Eq. \eqref{SystemForSolvingdAndfii0} outside this parameter space, but they are not physically reasonable. We have made this kind of contour analysis for every $G-V$ data set that we have studied, and found a single solution every time.

\section*{References}

\bibliography{aipsamp}

\end{document}